\documentclass[12pt]{article}
\usepackage{amsmath}
\usepackage{amssymb}
\begin{document}
\vskip 2cm
\begin{center}
{\sf {\Large   Self-Dual Chiral Boson: Augmented Superfield Approach}}

\vskip 3.0cm

{\sf D. Shukla$^{(a)}$, T. Bhanja$^{(a)}$, R. P. Malik$^{(a,b)}$}\\
$^{(a)}$ {\it Physics Department, Centre of Advanced Studies,}\\
{\it Banaras Hindu University, Varanasi - 221 005, (U.P.), India}\\

\vskip 0.1cm

{\bf and}\\

\vskip 0.1cm

$^{(b)}$ {\it DST Centre for Interdisciplinary Mathematical Sciences,}\\
{\it Faculty of Science, Banaras Hindu University, Varanasi - 221 005, India}\\
{\small {\sf {e-mails: dheerajkumarshukla@gmail.com; tapobroto.bhanja@gmail.com; rpmalik1995@gmail.com}}}

\end{center}

\vskip 2cm

\noindent
{\bf Abstract:} 
We exploit the standard tools and techniques of the augmented version of Bonora-Tonin (BT) 
superfield formalism to derive the off-shell nilpotent and absolutely anticommuting (anti-)BRST 
and (anti-)co-BRST symmetry transformations for the Becchi-Rouet-Stora-Tyutin (BRST)
 invariant Lagrangian density of a self-dual bosonic system. In the derivation of the 
full set of the above symmetry transformations, we invoke the (dual-)horizontality conditions, 
(anti-)BRST and (anti-)co-BRST invariant restrictions on the superfields that are defined on the
 (2, 2)-dimensional supermanifold. The latter is parameterized by the bosonic variable 
$x^\mu\,(\mu = 0,\, 1)$ and a pair of Grassmanian variables $\theta$ and $\bar\theta$ 
(with $\theta^2 = \bar\theta^2 = 0$ and $\theta\bar\theta + \bar\theta\theta = 0$). 
The dynamics of this system is such that, instead of the full (2, 2) dimensional superspace coordinates
 $(x^\mu, \theta, \bar\theta),$ we require {\it only} the specific (1, 2)-dimensional
 super-subspace variables $(t, \theta, \bar\theta)$ for its description. 
This is a {\it novel} observation in the context of superfield approach to BRST formalism.
 The application of the dual-horizontality condition, in the derivation of a set of proper 
(anti-)co-BRST symmetries, is {\it also} one of the {\it new} ingredients of our present endeavor where we have exploited
the augmented version of superfield formalism.\\



\newpage
\section{Introduction}

The model of a self-dual chiral boson is a widely  studied subject (see, e.g. [1-8]) because
it finds its use in the description of models of superstrings,  quantum Hall effect, W-gravities and
some of the two dimensional statistical systems. In a recent paper [9], this model has also been shown to 
be a field theoretical example for the Hodge theory because this two (1 + 1)-dimensional (2D) system
provides physical realizations of the de Rham cohomological operators of differential geometry in
the language of its symmetry properties (and corresponding conserved charges). The purpose
of our present investigation is to apply the augmented version of BT-superfield formalism [10-15]
to derive its proper nilpotent (anti-)BRST and (anti-)co-BRST symmetry transformations (which 
{\it together} lead to the derivation of a unique bosonic symmetry transformation [9]). These symmetries
collectively lead to the physically conceivable realizations of the cohomological operators.

The above superfield approach to BRST formalism is one of the geometrically intuitive methods to 
shed light on the abstract mathematical properties associated with the proper (anti-)BRST symmetries in the 
language of geometrical objects on the supermanifold. Usually, within the framework of this superfield 
approach, a given D-dimensional {\it ordinary} gauge theory is generalized onto a (D, 2)-dimensional 
supermanifold where the superfields are defined corresponding to the ordinary dynamical fields of a
given ordinary gauge theory. It is the (D, 2)-dimensional superspace coordinates\footnote {In the definition 
of $Z^M = (x^\mu, \theta, \bar\theta)$, the ordinary
coordinates $x^\mu$ (with $\mu = 0, 1, 2...D-1$) are the bosonic variables 
of the D-dimensional ordinary gauge theory and the pair of coordinates $\theta$ and $\bar\theta$ are the Grassmannian variables 
with their fermionic properties: $\theta^2 = \bar\theta^2 = 0, \, \theta{\bar\theta} + \bar\theta{\theta} = 0$. } 
$Z^M = (x^\mu ,\theta,\bar\theta )$  that characterize the 
(D, 2)-dimensional supermanifold. The gauge-invariant restrictions on the superfields lead to the derivation 
of (anti-)BRST symmetry transformations. On the other hand, the translational generators ($\partial_\theta,\,\partial_{\bar\theta}$), along 
the Grassmanian directions $(\theta,\bar{\theta})$, provide the geometrical basis for the (anti-)BRST symmetries 
and corresponding charges.

One of the decisive features of our present investigation is the key observation that, it is the 
{\it super-subspace} coordinates that are good enough to define the super exterior derivative and 
the supergauge connection in the description of the 2D self-dual chiral bosonic system within the 
framework of superfield formalism. In other words, even though we require (2, 2)-dimensional supermanifold to define the 
superfields (corresponding to the 2D ordinary fields of the self-dual chiral boson), 
the (dual-)horizontality conditions require (1, 2)-dimensional super-subspace variables
 for their description (see, e.g. Sec. 3 and Sec. 6). This peculiarity arises because of the fact 
that only {\it one} component of the ``gauge'' field (i.e. $ \lambda= \lambda _0 + \lambda _1$) is the {\it dynamical} 
as well as {\it propagating} field and its orthogonal counterpart $(\lambda _0 - \lambda _1)$ 
remains {\it inert}. This is a novel observation in the superfield approach to a given gauge theory
(when the latter is discussed within the framework of BRST formalism). Stated explicitly, we note that
our superfields, corresponding to the 2D ordinary fields of the chiral boson model, would be the function
of (2, 2)-dimensional superspace variables $Z^M = (x^\mu, \theta, \bar\theta)$. However, the super
exterior derivative $\tilde d$ and the super 1-form connection $\tilde A^{(1)}$ 
(that play important roles in restrictions on the appropriate superfields) would be defined in terms
of the (1, 2)-dimensional super-subspace variables $(t, \theta, \bar\theta)$ (cf. (19) below).

It is the above cited {\it novel} feature that has propelled us to pursue our present investigation 
 within the framework of our augmented version of BT-superfield formalism so that we could derive 
 the (anti-)BRST and (anti-)co-BRST symmetry transformations of this theory as well as 
provide the geometrical interpretations for them. 
One of the key signatures of a gauge theory is the existence of (anti-)BRST invariant Curci-Ferrari (CF) type 
restrictions [16] when this gauge theory is discussed within the framework of the superfield approach to BRST formalism. 
 For our 2D self-dual chiral bosonic system, we find that the CF-type conditions are {\it trivial} and they are found
to be (anti-)BRST as well as (anti-)co-BRST invariant, respectively (in the context of (anti-)BRST and (anti-)co-BRST
symmetries). Ultimately, in our present endeavor, we obtain the proper (i.e. nilpotent and absolutely anticommuting) 
(anti-)BRST and (anti-)co-BRST symmetries for the 2D system under consideration and provide 
 the geometrical basis for the symmetries and conserved charges. 
It is gratifying to state that we have checked that our proposal for the 
Hodge duality operation on the (1, 2)-dimensional super-subspace differentials (and their wedge products)
turns out to be correct and it leads to the derivation of proper (anti-)co-BRST symmetries
and (anti-)co-BRST invariant CF-type condition. The latter is a decisive feature of 
the BRST approach to a gauge theory.

As a background to our further discussions, we state a few facts about the 2D self-dual chiral bosonic
model. To start with, the covariant version of the self-dual chiral bosonic system
is {\it non-gauge invariant} but, after the inclusion of the Wess-Zumino term, it becomes a gauge invariant theory 
(see, e.g. Sec. 2). In other words, the inclusion of the Wess-Zumino term converts the second-class 
constraints of the original theory into their counterpart first-class constraints 
(thereby rendering the theory to be gauge-invariant). To corroborate the above statements, 
let us begin with the Lorentz covariant version of  Lagrangian density (see e.g. [17]) of the self-dual chiral boson
 in two (1 + 1)-dimensions of spacetime\footnote{We adopt here the convention
 and notations such that the 2D flat Minkowski spacetime is endowed with a flat metric with signatures 
(+1, -1) and the 2D Levi-Civita tensor is
chosen to satisfy: $\varepsilon_{\mu\nu}\, \varepsilon^{\mu\nu} = - 2\,!,\, \varepsilon_{\mu\nu} \,\varepsilon^{\mu\lambda} = - \,\delta _\mu ^\lambda $
with $ \varepsilon_{01} = +1 = -\varepsilon^{01}$. The overdot and prime on the fields, throughout the whole body of our text, always correspond to the partial derivatives w.r.t. time and space variables, respectively.}
\begin{eqnarray}
{\cal L}_{(0)} = \frac{1}{2} \partial_\mu \phi\,\partial^\mu \phi + \lambda_\mu (\varepsilon^{\mu\nu} + \eta^{\mu\nu})\, \partial_\nu \phi,
\end{eqnarray}
where the scalar field $\phi$ obeys the self duality condition (i.e. $\dot \phi = \phi^\prime$) due to the following Euler Lagrange (EL) 
equation of motion 
\begin{eqnarray}
(\varepsilon^{\mu\nu} + \eta^{\mu\nu})\, \partial_\nu \phi\, =\, 0\,\qquad \Longrightarrow\, \qquad \dot\phi\, =\, \phi'.
\end{eqnarray}
The other covariant form of the EL equation of motion:
\begin{eqnarray}
\Box \,\phi + (\varepsilon^{\mu\nu} + \eta^{\mu\nu})\, \partial_\nu \lambda_\mu = 0,
\end{eqnarray}
implies the masslessness condition $\Box\, \phi = 0$ and self-duality condition 
$\dot\lambda = \lambda'$ where $\lambda = \lambda_0 + \lambda_1$. Thus, it is clear that there are {\it two}
self-dual fields (i.e. $\dot\phi\, =\, \phi',\, \dot\lambda = \lambda' $ ) in our theory
at the classical level. The non-Lorentz covariant form of a {\it single} chiral boson was written in [5]
which was made Lorentz covariant form in a couple of very interesting papers [18,19]. It can be checked that,
in the component form (with $\eta_{\mu\nu} = $diag $(+1, -1) = \eta^{\mu\nu} $), we can have the following expression for this starting Lagrangian density (1), namely;
\begin{eqnarray}
{\cal L}_{(0)} = \frac{1}{2}\,(\dot\phi^2 - {\phi'}^2 ) + \lambda\,(\dot\phi - \phi^\prime),
\end{eqnarray}
which is normally used for the description of the self-dual chiral boson\footnote{At this stage,
this theory is non-gauge invariant. However, after the inclusion of the Wess-Zumino term, 
it becomes gauge invariant (see, e.g. Sec. 2 below).}. We re-emphasize that it is the
combination $\lambda = \lambda_0 + \lambda_1$ that participates in the description 
of the self-dual chiral boson but its orthogonal counterpart ($\lambda_0 - \lambda_1$) does not appear in the Lagrangian density.
In our further discussions, we shall focus $only$ on ($\lambda = \lambda_0 + \lambda_1$) as the
propagating  ``gauge" field (see, e.g. [17] for details) and shall completely ignore the non-propagating 
($\lambda_0 - \lambda_1$) component (see, Sec. 3 below) of the gauge-field.

Our present investigation is essential on the following counts. First, the self-dual chiral boson
is an interesting model which is required in many physical systems of importance. Thus, anything $novel$ 
about this model is interesting in itself. We observe some novelties in the application of the augmented 
version of superfield approach to BRST formalism for this system. As it turns out,  
we need only (1, 2)-dimensional super-subspace variables for its characterization.
Second, it was challenging for us to apply the theoretical arsenal of augmented version of superfield 
formalism to this system where only {\it one} component 
($\lambda = \lambda_0 + \lambda_1$) of the gauge field couples with the matter field and its
orthogonal component remains in the background. In our earlier applications of superfield
formalism (see, e.g. [12-15]) to (non-)Abelian gauge theories, we have {\it never} come 
across the {\it peculiarity} of the 2D chiral bosonic system under consideration. Finally, it was very important
for us to apply and check the sanctity of Hodge duality operation on the (2, 2)-dimensional
supermanifold in the determination of (anti-)co-BRST symmetries that are present in the theory.
As it turns out for our present theory, we need only (1, 2)-dimensional super-subspace variables 
($t,\theta ,\bar{\theta }$) for its description (and the Hodge duality is applied on these coordinates only).
We have accomplished all the above  goals in our present endeavor.

The contents of our present endeavor are organized as follows. In Sec. 2, we discuss the bare essentials of the
(anti-)BRST symmetry transformations (and corresponding conserved charges) for the self-dual chiral boson 
in the Lagrangian formulation. Our Sec. 3 is devoted to the derivation of the above (anti-)BRST transformations
using the augmented version of BT-superfield formalism. In Sec. 4, we deal with the proof of the (anti-)BRST
invariance of the super Lagrangian density and capture the nilpotency and absolute anticommutativity properties 
of the (anti-)BRST charges within the framework of our superfield approach. Our Sec. 5 contains a brief synopsis 
of the (anti-)co-BRST  symmetries (and corresponding charges) of our present theory. We derive these symmetries 
in our Sec. 6 using superfield formalism. Our Sec. 7 deals with the (anti-)co-BRST invariance of the Lagrangian density
 and nilpotency of the (anti-)co-BRST charges within the framework of our superfield approach. 
Finally, we make some concluding remarks and point out a few future directions 
for further investigations in Sec. 8.

In our Appendix A, we discuss the key differences between the 
nilpotent (anti-)BRST and (anti-)co-BRST symmetries and
our Appendices B and C are devoted to the explicit computations of (i) the 
dual-horizontality condition, and (ii) the auxiliary variable $\cal{B}$.

\noindent
\section{Preliminary: Nilpotent (Anti-)BRST Symmetries in Lagrangian Formulation}

Let us begin with the (2D) second-order Lagrangian density (${\cal L}_{b} $) 
for the description of a single self-dual chiral boson within the framework of BRST formalism. 
This effective Lagrangian density of the 2D chiral boson (with the inclusion of the Wess-Zumino term) 
is  as follows (see, e.g. [20], [9] for details)
\begin{eqnarray}
{\cal L}_b &=& \frac {\dot\phi^2}{2} - \frac{\dot v^2}{2} + \dot v\,(v' -\phi') + \lambda \,\bigl[\dot\phi -\dot v + v'- \phi' \bigr] 
- \frac{1}{2}\,(\phi' - v')^2\nonumber\\ &+& B\,(\dot\lambda - v - \phi)
+ \frac{B^2}{2} - i\, \dot{\bar C}\, \dot C + 2\, i \,\bar C\, C,\nonumber\\
&\equiv & \frac {\dot\phi^2}{2} - \frac{1}{2}\,\bigl[\dot v - (v' - \phi')\bigr]^2 + 
\lambda \,\bigl[\dot\phi -\dot v + v'- \phi' \bigr] + B\,(\dot\lambda - v - \phi)\nonumber\\
&+&  \frac{B^2}{2} - i\, \dot{\bar C}\, \dot C + 2\, i \,\bar C\, C,
\end{eqnarray}
where $\dot\phi = \partial \phi \slash \partial t,\, \dot v = \partial v \slash \partial t,
\, \dot{\bar C} = \partial \bar C \slash \partial t, \,\dot\lambda = \partial \lambda \slash \partial t,$
etc, are the generalized ``velocities" with respect to the evolution parameter $t$, $B$ is 
the Nakanashi- Lautrup auxiliary field, the fermionic ($ C^2 = \bar C^2 = 0,\, C\,\bar C + \bar C \, C = 0$) 
(anti-)ghost fields $(\bar C) C$ are required for the validity of unitarity in our theory and 
the notations $\phi' = (\partial \phi \slash \partial x),\, v' =\,(\partial v \slash \partial x) $ 
stand for the single space derivative on the 2D chiral boson field $\phi$ and Wess-Zumino (WZ) field $v $, 
respectively. We further note that:  $B\, (\dot\lambda - v- \phi) + \frac {1}{2}\,B^2 = - \frac{1}{2}\, (\dot\lambda -v - \phi)^2$ 
is the gauge-fixing term for the gauge field $\lambda(x)$ where $B = -\,(\dot\lambda - v - \phi)$.
The above Lagrangian density is the (anti-)BRST invariant generalization of (4) and it 
respects the following off-shell nilpotent and infinitesimal (anti-)BRST symmetry transformations $(s_{(a)b})$:
\begin{eqnarray}
&&s_{ab}\phi = - \bar C,\quad s_{ab} v = - \bar C, \quad s_{ab}\lambda = \dot{\bar C},\quad s_{ab} \bar C = 0,\quad
s_{ab} C = -\,i\,B,\quad s_{ab} B = 0,\nonumber\\ 
&&s_{b}\phi = - C,\,\,\quad s_{b} v = -  C,\,\, \quad s_{b}\lambda = \dot C,
\,\,\quad s_{b} \bar C = +\,i\,B,\quad s_{b} C = 0,\,\, \quad s_{b} B = 0.
\end{eqnarray}
In fact, it can be explicitly checked that the 
(anti-)BRST invariant Lagrangian density (5) transforms, under $s_{(a)b}$, to the total time derivatives as 
\begin{eqnarray}
s_{b}\,{\cal L}_b = \frac{\partial}{\partial t}\,\bigl[B\,\dot C\bigr], \qquad\qquad \qquad
s_{ab}\,{\cal L}_b = \frac{\partial}{\partial t}\,\bigl[B\,\dot {\bar C}\bigr].
\end{eqnarray}
Thus, the action integral $ S = \int dx \int dt \,{\cal L}_b $ remains invariant under the above (anti-)BRST 
transformations. Further, it is straightforward to note that the above (anti-)BRST symmetry transformations
(6) are {\it off-shell} nilpotent ($s_b^2 =  s_{ab}^2 = 0$) and absolutely anticommutating in nature ($ s_b\,s_{ab} + s_{ab}\,s_b = 0$)
without any use of the Euler-Lagrange (EL) equations of motion:
\begin{eqnarray}
\dot B &=& \dot\phi - \dot v + v' - {\phi}', \,\,\, B = v + \phi - \dot\lambda,\qquad \ddot C + 2 C = 0,\nonumber\\ 
- B &=& \ddot\phi + \dot\lambda - \dot v' - \lambda' - (\phi'' - v ''),\qquad\qquad \ddot{\bar C} + 2\,\bar C = 0,\nonumber\\
B &=& \ddot v - 2\,\dot v' + {\dot\phi}' + \dot\lambda - {\lambda}' -({\phi}'' - v''),
\end{eqnarray}
which emerge from the Lagrangian density (5). The off-shoot of the above equations is the observation that $\ddot B + 2 B = 0$ because 
$(- 2 B) = \ddot\phi - \ddot v + \dot v' - \dot\phi'$ which is equal to $\ddot B$ (as is evident from the first entry of the above equations: $\dot B = \dot\phi - \dot v + v' - \phi '$). We shall see, later on, that this relation is something like the
Curci-Ferrari type restriction [16] because it plays an important role in proving the anticommutativity property.

By exploiting the standard tricks and techniques of Noether's theorem, it can be checked that the conserved ($\dot Q_{(a)b} = 0$),
nilpotent ($ Q_{(a)b}^2 = 0 $) and absolutely anticommutating ($Q_b\,Q_{ab} + Q_{ab}\,Q_b = 0$) (anti-)BRST charges\footnote{ Our expressions for the off-shell nilpotent (anti-)BRST charges match with [20] but we differ slightly from the 
corresponding expressions quoted in [9] at the conceptual level.}:
\begin{eqnarray}
Q_b &=& \int dx \,\bigl[ B\dot C - ( \dot\phi - \dot v + v' - \phi' )\, C\bigr]
\,\,\equiv\,\,  \int dx\, \bigl[ B\,\dot C - \dot B \,C\bigr],\nonumber\\
Q_{ab} &=&  \int dx\, \bigl[ B\dot {\bar C} - ( \dot\phi - \dot v + v' - \phi' )\, \bar C\bigr]
\,\,\equiv \,\, \int dx \,\bigl[ B\,\dot {\bar C} - \dot B\, \bar C\bigr],
\end{eqnarray}
are the generators of the (anti-)BRST symmetry transformations (6). Using this property, it is straightforward to prove the following relationships:
\begin{eqnarray}
&& s_b Q_b = -i\, \{Q_b,\, Q_b\} = 0,\quad\qquad\qquad s_{b} Q_{ab} = -i\, \{Q_{ab},\, Q_{b}\} = 0\nonumber\\
&&s_{ab} Q_{ab} = -i\, \{Q_{ab},\, Q_{ab}\} = 0,\qquad\,\,\quad 
s_{ab} Q_{b} = -i\, \{Q_{b},\, Q_{ab}\} = 0,
\end{eqnarray}
which demonstrate the nilpotency and absolute anticommutativity property of the conserved (anti-)BRST charges $Q_{(a)b}$. In the proof of the conservation of the charges $Q_{(a)b}$, one has to exploit the off-shoot of the 
equation of motion (8) which shows that  $\ddot B + 2 B = 0$. In the forthcoming two subsequent sections, we shall capture these properties within the framework of our augmented version of BT-superfield approach.

We wrap up this section with the remark that there exists the global infinitesimal ghost-scale symmetry transformations\footnote{To be precise, the fields of our present theory transform as:
$\Psi\, \rightarrow e^{0\,\Omega }\,\Psi,\, C\,\rightarrow e^{1\Omega }\,C,\, \bar{C}\,\rightarrow e^{-1\Omega }\,\bar{C}$ where $\Omega$  is a global scale infinitesimal transformation parameter and the generic field $\Psi = \phi,\, v,\, \lambda$. The numerals in the
exponents denote the ghost number of the fields. We set
 $\Omega=\,1$ to obtain the simpler form of the infinitesimal version of these 
transformations as: ${s_g} C = C,\; {s_g} \bar{C}  = - \bar{C}, \;{s_g} \Psi = 0 $
and the algebra $s_g Q_b = - i [ Q_b, Q_g ] = + Q_b, s_g Q_{ab} = - i [ Q_{ab}, Q_g ] = - Q_{ab}$.}: 
$s_g\, C = C,\, s_g\, \bar C = -\,\bar C, \, s_g \Psi = 0,\, (\Psi = \phi,\, v,\, \lambda)$
in the theory which are generated by the Noether conserved ghost charge $Q_g = - i\,\int dx \bigl[(C\,\dot{\bar C} 
+ \bar C \,\dot C)\bigr]$. It is elementary to check that the ghost charge $Q_g$ satisfies the following algebra with the nilpotent and conserved (anti-)BRST charges:
\begin{eqnarray}
i\,\bigl[Q_g,\, Q_b\bigr] = Q_b, \qquad \qquad \quad i\,\bigl[Q_g,\, Q_{ab}\bigr] = -\,Q_{ab}.
\end{eqnarray}
The above algebra, together with the algebra quoted in (10), obeys the standard BRST algebra amongst the nilpotent
(anti-)BRST charges $Q_{(a)b}$ and the ghost charge $Q_g$, namely;
\begin{eqnarray}
&&\{Q_{(a)b},\,Q_{(a)b}\} = Q_{(a)b} ^2 = 0,\qquad\qquad \{Q_b,\,Q_{ab}\} = 0,\nonumber\\
&&i\,\bigl[Q_g,\, Q_b\bigr] = Q_b,\qquad\qquad\qquad\qquad\,\, i\,\bigl[Q_g,\, Q_{ab}\bigr] = -\,Q_{ab},
\end{eqnarray}
which establishes that the ghost number of the BRST charge is ($+1$) and that of the anti-BRST charge is ($-1$) 
and the (anti-)BRST charges are nilpotent of order two. We further note that one can derive the {\it on-shell} nilpotent (anti-)BRST symmetry transformations from (6) by replacing $ B = -\,(\dot\lambda -\, v -\, \phi)$. These symmetry transformations are:
\begin{eqnarray}
&&s_{ab}\phi = - \bar C,\quad s_{ab}\, v = - \bar C, \quad s_{ab}\lambda = \dot{\bar C},\quad
s_{ab}\, \bar C = 0,\quad s_{ab} \,C = +\,i\,(\dot\lambda -\, v -\, \phi),\nonumber\\
&&s_{b}\,\phi = - C,\qquad s_{b}\, v = -  C, \quad s_{b}\,\lambda = \dot C,
\quad s_{b}\, C = 0,\quad s_{b}\, \bar C = -\,i\,(\dot\lambda -\, v -\, \phi).
\end{eqnarray}
For the sake of brevity, we have adopted the same notations for the off-shell as well as on-shell (anti-)BRST symmetry transformations.
We further observe that:
\begin{eqnarray}
s_{ab}^2 \,C = -\,i\,\bigl[\ \ddot{\bar C} + 2\, \bar C \,\bigr],\qquad \qquad s_{b}^2 \,\bar C = +\,i\,\bigl[\, \ddot C + 2\, C \,\bigr]. 
\end{eqnarray}
The r.h.s of the above transformations are zero on the {\it on-shell} where the equations of motion: 
$\ddot{\bar C} + 2\, \bar C = 0,\quad \ddot C + 2\, C = 0   $ are valid. Similarly, we point out that the absolute anticommutativity
property of (13) is satisfied if and only if the equations (8) are taken into account.
The above on-shell nilpotent symmetry transformations are {\it true} symmetry transformations  for the 
following Lagrangian density (without presence of the Nakanishi-Lautrup field $B$):
\begin{eqnarray}
{\cal L}_{b}^{(0)}&=& \frac {\dot\phi^2}{2} - \frac{\dot v^2}{2} + \dot v\,(v' -\phi') 
+ \lambda \,\bigl[\dot\phi -\dot v + v'- \phi' \bigr] 
- \frac{1}{2}\,(\phi' - v')^2\nonumber\\ &-& \frac{1}{2}\,(\dot\lambda - v - \phi)^{2}  - i\, \dot{\bar C}\, \dot C
+ 2\, i \,\bar C\, C,
\end{eqnarray} 
because we have the following transformations:
\begin{eqnarray}
s_b\,{\cal L}_{b}^{(0)} = -\, \frac{\partial}{\partial t}\,\bigl[\,(\dot\lambda - v - \phi)\,\dot C],\qquad\quad 
s_{ab}\,{\cal L}_{b}^{(0)}  = -\, \frac{\partial}{\partial t}\,\bigl[\,(\dot\lambda - v - \phi)\,\dot {\bar C}].
\end{eqnarray}
The above observations demonstrate that the action integral $ S = \int d^2 x  \,{\cal L}_{b}^{(0)} $ remains 
invariant under the on-shell nilpotent symmetry transformations (13). The expressions for the on-shell nilpotent
conserved charges $Q_{(a)b}$ can be derived from (9) by the replacements: $B = -\,(\dot{\lambda }- v - \phi) 
\equiv -\,\frac{1}{2}\, [(\ddot{\phi} - \ddot{v}) - (\dot{\phi}' - \dot{v}') ] $  and 
$\dot{B}= (\dot{\phi } -\dot {v}) - (\phi^\prime - v^\prime) $. 
The stage is now set to discuss about the original
theory (1) at the quantum level. In this connection, we note that our original theory (1) has been modified due
to presence of the WZ field $v$. Within the framework of BRST formalism, 
the physical state $ |phys > $ of the theory is defined as $ Q_{(a)b}\,| phys >\, =\, 0$ at the 
quantum level. Using expressions for $ Q_{(a)b}$ from (9), it is clear that  
$ B \,|phys>\, =\, 0,\,\, \dot{B}\,| phys > \,= \,0 $ due to the fact that the (anti-)ghost fields are 
decoupled from the rest of the theory and they themselves do not produce any physically meaningful constraints
on the theory. 
From the EL equations of motion (8),
it is clear that the above conditions imply that $[(\dot{\phi} - \dot{v}) - (\phi' - v')]\,| phys >\, =\, 0$
and $ [(\ddot{\phi} - \ddot{v}) - (\dot{\phi}' - \dot{v}') ]\,| phys >\, =\, 0 $.
With the redefinition: $ \tilde{\phi} = (\phi - v) $, it is evident that, ultimately, we have the conditions
on the physical state as: $ (\dot{\tilde{\phi}} - \tilde{\phi}')\,| phys > \,= \,0 $
and $ \frac{d}{dt}\,(\dot{\tilde{\phi}} - \tilde{\phi}')\,| phys > \,= \,0 $
which imply the self-duality condition and its time-evolution invariance
\footnote{The WZ field $ v $ is introduced by hand 
(i.e. $ \phi \longrightarrow \phi - v,\,\, \lambda\longrightarrow \lambda + \dot{v} $) 
to convert the Lagrangian density (4) into (5) (modulo the gauge-fixing and FP ghost terms). 
The redefinition of $ \phi $ gets rid of WZ field $ v $ from our theory at the quantum level.}.
Thus, at the quantum level, in reality, there exists only a single self-dual chiral boson $\phi$ and the duality condition
$\dot \lambda = \lambda^\prime$ does {\it not} appear at the quantum level in any physically meaningful limit.

We shall concentrate on the Lagrangian density (15) 
for our further discussions on the (anti-)dual-BRST symmetry transformations (Sec. 5).

\noindent
\section {(Anti-)BRST Symmetries: Superfield Formalism}

To derive the (anti-)BRST symmetry transformations (6) within the framework of the augmented superfield formalism, 
first of all, we generalize the basic variables of the 2D theory onto the (2, 2)-dimensional supermanifold as follows:
\begin{eqnarray}
&&\phi(x) \longrightarrow \tilde\Phi(x, \theta, \bar\theta),\qquad  v(x) \longrightarrow \tilde V(x,\theta,\bar\theta ),\qquad \lambda (x) \longrightarrow \tilde\lambda(x, \theta, \bar\theta)\nonumber\\
&&C(x) \longrightarrow F(x, \theta, \bar\theta),\qquad \bar C(x) \longrightarrow \bar F(x, \theta, \bar\theta),\nonumber\\
\end{eqnarray}
where $ Z^M =(x^{\mu}, \theta, \bar\theta)$ is the superspace coordinate that parameterizes the above (2, 2)-dimensional 
superfield where, out of the spacetime bosonic pair $(x, t)$, {\it t} is the bosonic (evolution) parameter of the 
theory and the set $(\theta,\, \bar\theta)$ is a pair of Grassmanian variables  
(with $\theta^2 = \bar\theta^2 = 0$ and $\theta\bar\theta + \bar\theta\theta = 0$).
The gauge variable of the 2D effective theory is $\lambda (x)$ (i.e. $\lambda(x) = \lambda_0 (x) + \lambda_1(x)$) which couples
with the chiral bosonic field $\phi (x)$ and its orthogonal counterpart $(\lambda_0 (x) - \lambda_1 (x))$ remains {\it inert} in our
whole discussion. Thus, we define the appropriate 1-form gauge connection as 
\begin{eqnarray}
A^{(1)} &=& dx^\mu\; \lambda_\mu \equiv \frac{1}{2} \;(d x^0 + d x^1) \; (\lambda_0  + \lambda_1)
+ \frac{1}{2} \;(d x^0 - d x^1)\; (\lambda_0 - \lambda_1) \nonumber\\
&\equiv& d x^{+}\; (\lambda_0 + \lambda_1) + dx^{-} \; (\lambda_0 - \lambda_1) \equiv dt \; \lambda,
\end{eqnarray}
where we have ignored the $dx^{-}$ component because of our preceding discussions and have identified $d x^{+}$
component with the differential $dt$ that corresponds to the evolution parameter of our theory.
The exterior derivative $ d = dt\, \partial_t$ (with $d^2 = 0$) has been taken
in our theory because of our above arguments. It is evident that $d\,A^{(1)} = 0 $ due to $ (dt \wedge dt) = 0$.
These 1-forms of geometrical interest can be generalized onto our chosen (2, 2)-dimensional supermanifold as:
\begin{eqnarray}
&&d\longrightarrow \tilde d = dZ^M\partial_ M \equiv  dt\,\partial _ t + d\theta\, \partial _\theta 
+ d\bar\theta\, \partial _{\bar\theta} ,\,\, \tilde d ^2 =0, \nonumber\\
&&A^{(1)} \longrightarrow \tilde A^{(1)} = dZ^M A_M
\equiv dt\,\tilde\lambda(x, \theta, \bar\theta) + d\theta\, \bar F(x, \theta, \bar\theta) + d\bar\theta\, F(x, \theta, \bar\theta),
\end{eqnarray}
where $\partial_M = {\partial}\slash {\partial Z^{M}} \equiv (\partial_t,\, \partial_\theta,\, \partial_{\bar\theta})$ 
corresponds to the set of super-subspace partial derivatives corresponding to the (1, 2)-dimensional super-subspace coordinates
$(t,\theta ,\bar\theta )$ and $A_M \equiv (\tilde\lambda\,(x, \theta, \bar\theta),\, F(x,\, \theta,\, \bar\theta),\, \bar F(x, \theta, \bar\theta) )$ 
is the vector supermultiplet of superfields defined on the (2, 2)-dimensional supermanifold that is
parametrized by $(x^+, x^{-}, \theta, \bar\theta) = (t, x, \theta, \bar\theta)$ in explicit components of the spacetime
variables $(t, x)$ and Grassmannian variables $(\theta, \bar\theta)$.
We re-emphasize that, the dynamics of our problem is such that we require only the 
superspace coordinates $(t, \theta, \bar\theta)$.
This is precisely the reason that we have the above kind of generalizations (19)
which are confined to (1, 2)-dimensional super-submanifold only.

The superfields (17) can be expanded along the full Grassmanian directions 
(i.e., $ 1,\, \theta,\,\bar\theta,\,\theta\,\bar\theta $) of the full  (2, 2)-dimensional supermanifold as:
\begin{eqnarray}
&&\tilde V (x, \theta, \bar\theta ) = v(x) + i\, \theta \,\bar f_2(x) +i\, \bar\theta\, f_2(x) + i\, \theta\,\bar\theta\, b_2(x), \nonumber\\
&& \tilde\Phi (x, \theta, \bar\theta) = \phi(x) + i\, \theta \,\bar f_1(x) + i\, \bar\theta\, f_1 (x) + i\, \theta\,\bar\theta\, b_1(x), \nonumber\\
&& F(x, \theta, \bar\theta ) = C(x) + i\, \theta\,\bar B_1 (x) + i\, \bar\theta\, B_1(x) + i\, \theta\,\bar \theta\, s(x), \nonumber\\
&& \bar F(x, \theta, \bar\theta ) = \bar C(x) + i\, \theta\, \bar B_2(x) + i\, \bar\theta\, B_2(x) + i\, \theta\,\bar\theta\, \bar s(x), \nonumber\\
&& \tilde\lambda (x, \theta, \bar\theta) = \lambda(x) + \theta\, \bar R(x) + \bar\theta \,R(x) + i\, \theta\, \bar\theta\, S(x),
\end{eqnarray}
where the secondary variables ($ R, \bar R, s,  \bar s , f_1, \bar f_1, f_2, \bar f_2  $)
on the r.h.s are fermionic and ($ S, B_1, \bar B_1, B_2, \bar B_2, b_1, b_2 $) are the bosonic secondary fields, respectively.   
It is obvious that we have the basic fields of the effective theory as $(\lambda, \phi, v, C, \bar C)$ which
are the limiting cases when we put $\theta = \bar\theta = 0$. We shall derive the 
secondary fields in terms of the basic fields (as well as auxiliary fields) of the effective 2D 
theory by exploiting some physically motivated restrictions.
First of all, let us exploit the horizontality condition (HC) which requires that 
\begin{eqnarray}
\tilde d \tilde A^{(1)} = d A^{(1)} = 0,
\end{eqnarray}
where the explicit form of the l.h.s (i.e. $\tilde d \tilde A^{(1)}$) is 
\begin{eqnarray}
&&(dt \wedge d \theta )\bigl[\partial_t \bar F -\, \partial_{\theta} {\tilde\lambda} \bigr] + 
(dt \wedge d {\bar\theta} )\bigl[\partial_{t} F -\, {\partial _{\bar\theta}}\tilde\lambda \bigr] 
- (d\theta \wedge d\bar\theta )\bigl[\partial_{\theta} F -\, \partial_{\bar\theta} \bar F \bigr]\nonumber\\
&&-\, (d\theta \wedge d\theta )\, \partial_{\theta} \,\bar F -\, (d {\bar\theta} \wedge d{\bar\theta )}\, \partial_{\bar\theta}\, F.
\end{eqnarray}
Setting the coefficients of the differentials $(dt \wedge d\theta ), (dt \wedge d\bar\theta ),
(d\theta \wedge d\theta ), (d\theta \wedge d\bar\theta ), (d\bar\theta \wedge d\bar\theta )$
equal to zero, we get the following relationships amongst the secondary fields
and basic (as well as auxiliary) fields of our present theory, namely;
\begin{eqnarray}
&&\bar B_1 + B_2 = 0,\qquad \quad B_1 = \bar B_2 = 0,\qquad \quad \; \;s = 0,\nonumber\\
&& R = \dot C,\qquad \quad \bar R = \dot{\bar C},\qquad \quad S = \dot B,\qquad \,\, \bar s = 0,
\end{eqnarray}
where we have used the expansions from (20). It is essential to point out that when we put the coefficients of
($d\theta \wedge d\bar\theta$) equal to zero, we obtain the Curci-Ferrari type of restriction $\bar B_1 + B_2 = 0$ which is 
{\it trivial} for our simple theory.
If we choose $B_2 = B $ and $ \bar B_1 = - B$, we obtain the following explicit expansions for the superfields (20), namely; 
\begin{eqnarray}
\tilde\lambda^{(h)}(x, \theta, \bar\theta) &=& \lambda + \theta\, \dot{\bar C} + \bar\theta\, \dot C + \theta\,\bar\theta\, (i\,\dot B)\nonumber\\
& \equiv & \lambda + \theta\, (s_{ab}\, \lambda ) + \bar\theta\, (s_b\, \lambda )  + \theta\,\bar\theta\, (s_b\, s_{ab}\, \lambda ),\nonumber\\
F^{(h)}(x, \theta, \bar\theta) &=& C + \theta\, (-i\,B) + \bar\theta\, (0) + \theta \,\bar\theta \,(0) \nonumber\\
& \equiv &  C + \theta \,(s_{ab} C) +\bar\theta\, s_b \,(C) +  \theta\,\bar\theta \,(s_b\, s_{ab}\, C), \nonumber\\
{\bar F}^{(h)}(x, \theta, \bar\theta) &=& \bar C + \theta\, (0) + \bar\theta\, (i\,B) + \theta\, \bar\theta\, (0) \nonumber\\
& \equiv & \bar C + \theta\, (s_{ab}\, \bar C) +\bar\theta\, (s_b \bar C) +  \theta\,\bar\theta \,(s_b\, s_{ab}\, \bar C),  
\end{eqnarray}
where the superscript ($h$) stands for the expansion of the superfields after the application of HC. 
A close look at (24) establishes that we have already derived the following: 
\begin{eqnarray}
&&s_b\, \lambda = \dot C, \qquad\qquad s_{ab}\, \lambda =\, \dot{\bar C}, \,\;\quad\qquad\qquad s_b \,s_{ab}\,\lambda = i\,\dot B,\nonumber\\
&&s_b\, C = 0, \qquad\qquad s_{ab}\, C = -i\, B, \qquad\qquad s_b \,s_{ab}\, C = 0, \nonumber\\
&&s_b\, \bar C = i \,B, \,\quad\qquad s_{ab}\, \bar C = 0,\, \,\;\quad\qquad\qquad s_b\, s_{ab}\,\bar C = 0.
\end{eqnarray}
The requirement of nilpotency, in the above, implies that $s_b \,B = 0$ and $s_{ab}\, B = 0$.

To obtain the off-shell nilpotent and absolutely anticommuting (anti-)BRST symmetry transformations 
for $\phi (t)$ and $v(t)$ variables, we have to exploit the key ideas of the augmented version of BT-superfield 
formalism [12-15] where we demand that all the (anti-)BRST invariant quantities should remain independent of the ``soul" coordinates 
(i.e., $\theta $ and $\bar\theta$ ) when they are generalized onto the supermanifold. Thus, we invoke 
the following (anti-)BRST invariant restrictions on the (super)fields, namely;
\begin{eqnarray}
\tilde\lambda ^{(h)} (x, \theta, \bar\theta) + \dot{\tilde\Phi}(x, \theta, \bar\theta) = \lambda (x) + \dot \phi (x), \nonumber\\
\tilde\lambda ^{(h)}(x, \theta, \bar\theta) + \dot{\tilde V}(x, \theta, \bar\theta) = \lambda (x) + \dot v (x), 
\end{eqnarray}
which have been taken due to our observations that $s_{(a)b} \bigl[\lambda (x) + \dot \phi (x) \bigl] = 0$ 
and $s_{(a)b} \bigl[\lambda (x) + \dot v (x) \bigl] = 0$. We note that the (anti-)BRST invariant 
restrictions (26) and HC are intertwined together because the expansion for $\tilde\lambda ^{(h)}(x, \theta, \bar\theta)$ 
has to be taken from (24) which has been derived after the application of the HC 
(i.e. $\tilde d \tilde A^{(1)} = d A^{(1)} = 0$) in our theory.

Explicit substitution of (20) and (24) into (26) leads to the following relationships:
\begin{eqnarray}
f_1 = iC,\qquad \bar f_1 = i \bar C, \qquad b_1 = -\,B, \nonumber\\
f_2 = iC,\qquad \bar f_2 = i \bar C, \qquad b_2 = -\,B.
\end{eqnarray}
Thus, we have already obtained the expressions for the secondary fields in terms of the basic fields and auxiliary field ($B$). 
Finally, we have the following expansions: 
\begin{eqnarray}
\tilde\Phi ^{(b)}(x, \theta, \bar\theta) &=& \phi + \theta\, (-\,\bar C)  + \bar\theta\, (-\, C) + \theta\,\bar\theta\, (-\,i\,B)\nonumber\\
&\equiv& \phi + \theta \,(s_{ab}\, \phi ) + \bar\theta\, (s_b \,\phi )  + \theta\,\bar\theta \,(s_b\, s_{ab}\, \phi ),\nonumber\\
\tilde V^{(b)}(x, \theta, \bar\theta) &=& v + \theta\, (-\,\bar C)  + \bar\theta\, (-\,C) + \theta\,\bar\theta \,(-\,i\,B)\nonumber\\
&\equiv& v + \theta \,(s_{ab}\, v ) + \bar\theta\, (s_b \,v )  + \theta\,\bar\theta\, (s_b \,s_{ab}\, v ),
\end{eqnarray}
where the superscript ($b$) stands for the expansions of the supervariables after the application
of the nilpotent (anti-)BRST invariant restrictions.

A close look at the above expansions yield the following (anti-)BRST symmetry transformations for the fields $\phi (t)$ and $v(t)$:
\begin{eqnarray}
&&s_b\, \phi = - \,C, \quad\quad s_{ab} \,\phi =\, - \,{\bar C}, \quad\quad s_b \,s_{ab}\,\phi = -\,i\, B,\nonumber\\
&&s_b \,v = -\,C, \quad\quad s_{ab}\, v = -\, \bar C, \,\quad\quad s_b\, s_{ab} \,v = -\,i\, B. 
\end{eqnarray}
We draw the conclusion that we have already derived the full set of off-shell nilpotent and
absolutely anticommuting (anti-)BRST symmetry transformations
for the chiral bosonic system within the 
framework of the superfield approach to BRST formalism (cf. (25),(29)). We would like to close this 
section with the remark that there is a mapping between (anti-)BRST symmetry transformations and 
the translational generators along the Grassmanian directions of the full (2, 2)-dimensional supermanifold 
as can be seen from the following relationships:
\begin{eqnarray}
\frac{\partial}{\partial\theta} \Omega ^{(h,b)} (x, \theta, \bar\theta ) \mid _{\bar\theta = 0}\, = s_{ab}\, \omega (x),\quad\quad
\frac{\partial}{\partial\bar\theta} \Omega ^{(h,b)} (x, \theta, \bar\theta ) \mid _{\theta = 0} \,= s_{b}\, \omega (x),
\end{eqnarray}
where $\Omega ^{(h,b)} (x, \theta, \bar\theta )$ is the generic superfield (obtained after the 
application of the HC and (anti-)BRST invariant restrictions) and $\omega(x)$ is the generic field
of the effective 2D system. It is evident that the nilpotency and 
absolute anticommutativity properties of the (anti-)BRST  transformations $s_{(a)b}$ are intimately connected 
with such kinds of properties associated with the  Grassmanian translational generators 
$\partial_\theta$ and $\partial_{\bar\theta}$ (i.e. ${\partial_\theta}^2\, = {\partial_{\bar\theta}}^2 \,= 0 , 
\quad \partial_{\theta}\, \partial_{\bar\theta} + \partial_{\bar\theta}\, \partial_{\theta} \,= 0$).

\noindent
\section{(Anti-)BRST Invariance, Nilpotency and Anticommutativity: Superfield Approach}

In this section, we provide the geometrical basis for the existence of the nilpotency and absolute anticommutativity 
properties of the (anti-)BRST charges and the (anti-)BRST invariance of the Lagrangian density (5) (and the corresponding action) 
within the framework of the superfield formalism. Let us begin with the (anti-)BRST invariance of the action integral
$ S = \int dx \int dt \,{\cal L}_b $ that has been illustrated in equation (7). Having obtained the expressions in (24) and (28), 
it can be readily seen that the Lagrangian density (5) can be generalized onto the (2, 2)-dimensional supermanifold in the following manner:
\begin{eqnarray}
{\cal L}_b &\longrightarrow &  \tilde {\cal L}_b =\,\frac{1}{2} \bigl[\,\dot{\tilde\Phi}^{(b)}\,\dot{\tilde\Phi}^{(b)} 
- \dot{\tilde V}^{(b)}\,\dot{\tilde V}^{(b)}\bigr] 
+ \dot{\tilde V}^{(b)}\,\bigl[{\tilde V'^{(b)}} - {\tilde\Phi'^{(b)}}\bigr]
 - \frac{1}{2}\,\bigl[ {\tilde\Phi'^{(b)}} - {\tilde V'^{(b)}}\bigr]^2\nonumber\\
&+& \tilde\lambda^{(h)} \bigl[\dot{\tilde\Phi}^{(b)} - \dot{\tilde V}^{(b)} 
+ {\tilde V'^{(b)}} - {\tilde\Phi'^{(b)}}\bigr]
+ B (x) \,\bigl[ \dot{\tilde\lambda}^{(h)} - {\tilde V^{(b)}} - \tilde\Phi^{(b)} \bigl] \,+ \,\frac{B^2 (x)}{2}\nonumber\\
&-& \,i\,\dot{\bar F}^{(h)} \,\dot{F}^{(h)} + 2\,i\, {\bar F}^{(h)}\, {F}^{(h)},
\end{eqnarray}
where it is elementary to check that, we have:
\begin{eqnarray} 
\dot{\tilde\Phi}^{(b)}(x, \theta, \bar\theta ) -\dot{\tilde V}^{(b)} (x, \theta, \bar\theta )
- \tilde {V}'^{(b)}(x, \theta, \bar\theta) -\,\, \tilde{\Phi}'^{(b)}(x, \theta, \bar\theta )
&=& \dot\phi (x) - \dot v (x) + v' (x) -\, \phi' (x), \nonumber\\
\tilde {V}'^{(b)}(x, \theta, \bar\theta ) - \tilde{\Phi}'^{(b)}(x, \theta, \bar\theta ) &=& v'(x) - \,\phi'(x)
\end{eqnarray}
Now, it is straightforward to observe that 
\begin{eqnarray}
\frac{\partial}{\partial \bar\theta}\,\tilde {\cal L}_b|_{\,{\theta = 0}} = \frac{\partial}{\partial t}\,\bigl[B\,\dot C \bigr], 
\qquad\qquad \frac{\partial}{\partial \theta}\,\tilde {\cal L}_b|_{\,{\bar\theta = 0}} = \frac{\partial}{\partial t}\,\bigl[B\,\dot {\bar C} \bigr],
\end{eqnarray}
which matches with the equation (7) obtained in Sec. 2. It is worthwhile to mention that, in the above computation,
we have first calculated all the $\theta$ and $\bar\theta$ dependent terms separately  
and, then, have taken the derivatives with respect to $\bar\theta$ and $\theta$.

In view of the mapping (30), it is self-evident that we have obtained the following:
\begin{eqnarray}
\frac{\partial}{\partial \bar\theta}\,\tilde {\cal L}_b|_{\,{\theta = 0}} \longleftrightarrow s_b\,{\cal L}_b, \qquad\qquad
\frac{\partial}{\partial \theta}\,\tilde {\cal L}_b|_{\,{\bar\theta = 0}} \longleftrightarrow s_{ab}\,{\cal L}_b.
\end{eqnarray}
Geometrically, we have the following interpretation. The BRST and  anti-BRST invariance of the Lagrangian density 
in the ordinary space can be explained by the shift transformations of the super Lagrangian density along the
 $\bar\theta$ and $\theta$ directions of the supermanifold, respectively, 
such that the outcome is a total spacetime derivative in the ordinary
space. In other words, the physical (anti-)BRST invariance can 
be mathematically stated as the observation that the operation
of the derivative w.r.t. Grassmanian variables (i.e. $\partial_{\bar\theta}$ and 
$\partial_{\theta}$) on the super Lagrangian density (31)
produces the total time derivatives in the ordinary space.

Now we dwell on the proof of the nilpotency and anticommutativity of the (anti-)BRST charges within the framework of 
superfield formulation. First of all, it can be clearly checked that the (anti-)BRST charges can be expressed in terms of the
expansions (24) and (28) in the following {\it two} equivalent forms  
\begin{eqnarray}
Q_{ab} &=& \int dx\,\Bigl [\frac{\partial}{\partial\theta}\,\bigl[i\,(\bar F ^{(h)} \, \dot{F}^{(h)} 
+  F^{(h)} \dot{\bar F}^{(h)})\bigr]|\,_{\bar\theta = 0} \;\Bigr ]\nonumber\\
&\equiv& \int dx\, \Bigl [\;\int d\theta \,\bigl[i\,(\bar F ^{(h)} \, \dot{F}^{(h)} + F^{(h)} \dot{\bar F}^{(h)}) \bigr]|\,_{{\bar\theta = 0}}
\;\Bigr ],\nonumber\\
Q_{b} &=& \int dx\,\Bigl [\frac{\partial}{\partial{\bar\theta}}\bigl[-\,i\,( \bar F^{(h)}
 \, \dot{F}^{(h)} + \, F^{(h)} \dot{\bar F}^{(h)})\bigr]|\,_{\theta = 0}
\; \Bigr ] \nonumber\\
&\equiv& \int dx\,\Bigl [\; \int d{\bar\theta}\, \bigl[ -\,i\,(\bar F ^{(h)} \, \dot{F}^{(h)} 
+ F^{(h)} \dot{\bar F}^{(h)} )\bigr]|\,_{{\theta = 0}}\; \Bigr ].
\end{eqnarray}
In view of the mappings (30), the above equation (35) can be translated into the ordinary space in terms of the (anti-)BRST symmetry
transformations as:
\begin{eqnarray}
Q_{ab} = \int dx\, s_{ab}\,\Bigl[ i\,(\bar C\, \dot C + C \dot{\bar C})\Bigr],\qquad\qquad
Q_{b} = \int dx\, s_{b}\,
\Bigl[ -\,i\,(\bar C\, \dot C + C \dot{\bar C})\Bigr].
\end{eqnarray}
It is now elementary to prove that $\partial_{\theta}\, Q_{ab} = 0 $, $\partial_{\bar\theta}\, Q_{b} = 0$
and $s_{ab}\, Q_{ab} = 0$, $s_{b}\, Q_{b} = 0$ (cf. (35), (36)). These relationships provide a connection between
the nilpotency property associated with ${\partial_{\theta}}^2 = {\partial_{\bar\theta}}^2 = 0 $ and $s_{(a)b} ^2 = 0$. Furthermore, 
the definition of generators in terms of conserved charges $ s_b\, Q_b = -\,i\, \{ Q_b, \,Q_b\} = 0$ and 
$ s_{ab}\, Q_{ab} = -\,i\, \{ Q_{ab}, \,Q_{ab}\} = 0$ prove the nilpotency of the 
charges $Q_{(a)b}$ in a subtle manner.

In view of expansions in (24) and (28),
there are other ways to express the (anti-)BRST charges. These are explicitly written as follows:
\begin{eqnarray}
&&Q_{ab} =\int dx \,\Bigl[\frac{\partial}{\partial{\bar\theta}} \,\frac{\partial}{\partial \theta} 
\bigl (i \tilde\lambda^{(h)}\, \bar F^{(h)} \bigr )\Bigr],\qquad\quad\quad 
 Q_{b} = \int dx\,\Bigl[\frac{\partial}{\partial{\bar\theta}} \,\frac{\partial}{\partial \theta}\, 
\bigl (i \tilde\lambda^{(h)}\,\, F^{(h)} \bigr )\Bigr],\nonumber\\
&& Q_{ab} =\int dx \Bigl[\frac{\partial}{\partial{\bar\theta}} \,
\bigl (i\, \dot{\bar F}^{(h)}\, \bar F^{(h)} \bigr )\Bigr],\qquad\qquad\,\quad  
Q_{b} =\int dx\,\Bigl[\frac{\partial}{\partial{\theta}} 
\,\bigl (-\,i\, \dot{F}^{(h)}\, F^{(h)} \bigr )\Bigr],\nonumber\\
&& Q_{ab} =\int dx\, \Bigl[\frac{\partial}{\partial{\bar\theta}} \,
\frac{\partial}{\partial \theta} \bigl (-\,i \dot{\tilde\Phi}^{(b)}\, \bar F^{(h)} \bigr )\Bigr],\qquad \quad 
Q_{b} = \int dx\,\Bigl[\frac{\partial}{\partial{\bar\theta}}\,\frac{\partial}{\partial \theta}
\bigl  (-\,i \dot{\tilde {\Phi}}^{(b)}\, F^{(h)} \bigr )\Bigr],\nonumber\\
&& Q_{ab} =\int dx \,\Bigl[\frac{\partial}{\partial{\bar\theta}} \,\frac{\partial}{\partial \theta} 
\bigl (-\,i \dot{\tilde V}^{(b)}\, \bar F^{(h)} \bigr ) \Bigr], \qquad \quad 
Q_{b} =\int dx \,\Bigl[\frac{\partial}{\partial{\bar\theta}} \,\frac{\partial}{\partial \theta}
\bigl  (-\,i \dot{\tilde {V}}^{(b)}\, F^{(h)} \bigr )\Bigr],
\end{eqnarray}
which can be translated into the ordinary space in terms of symmetries $s_{(a)b}$ as follows:
\begin{eqnarray}
&& Q_{ab} =\int dx \,\Bigl[s_{b}\,s_{ab} \, ( i\,\lambda\, \bar C)\Bigr], \qquad\quad \quad 
Q_{b} =\int dx\, \Bigl[s_{b}\,s_{ab} \, ( i\,\lambda \,C)\Bigr],\nonumber\\ 
&&Q_{ab} =\int dx\,\Bigl[ s_{b} \, ( i\,\dot{\bar C} \,\bar C)\Bigr], \qquad\qquad \quad 
Q_{b} = \int dx \,\Bigl[s_{ab} \, (- \,i\,\dot C \, C)\Bigr],\nonumber\\
&& Q_{ab} =\int dx\,\Bigl[ s_{b}\, s_{ab} \, (- i\,\dot{\phi}\, \bar C)\Bigr],\qquad\quad 
Q_{b} = \int dx\, \Bigl[s_{b}\, s_{ab} \, (- i\,\dot{\phi}\, C)\Bigr],\nonumber\\
&& Q_{ab} = \int dx\,\Bigl[ s_{b}\, s_{ab} \, (- i\,\dot v\, \bar C)\Bigr],\qquad\quad 
Q_b =  \int dx\,\Bigl[s_{b}\, s_{ab} \, (- i\,\dot v\, \bar C)\Bigr]. \nonumber\\ 
\end{eqnarray}
The above expressions prove the nilpotency and absolute anticommutativity in a very simple manner. This is due to
the fact that   ${\partial_{\theta}}^2 = {\partial_{\bar\theta}}^2 = 0 $,
$\partial_{\theta} \,\partial_{\bar\theta}\, + \,\partial_{\bar\theta} \,\,\partial_{\theta} = 0$.
For instance, it can be seen that $Q_{b} = \int dx\, s_{ab}\, (-\,i\,\dot C\,C)$ implies that $s_{ab}\,Q_{b} = 0$ due to 
the nilpotency of $s_{ab}$. However, if we exploit the definition of the generator $ s_{ab}\, Q_b = -\,i\,\{Q_b,\, Q_{ab}\}$, 
we observe that the nilpotency of $s_{ab}$ (i.e. $s_{ab}^2 = 0$) is also connected with the absolute 
anticommutativity of the (anti-)BRST charges (i.e. $\{Q_b,\,Q_{ab}\} = 0$). 
Similar explanations can be provided for all the other expressions for the (anti-)BRST charges which are listed in (38).

We close this section with the remarks that the nilpotency and absolute anticommutativity of the  
(anti-)BRST symmetry transformations ($ s_{(a)b}^2 = 0,\, s_b\,s_{ab} +  s_{ab}\,s_b = 0 $) 
as well as the (anti-)BRST charges ($ Q_{(a)b}^2 = 0,\, Q_b\,Q_{ab} +  Q_{ab}\,Q_b = 0$) owe their origin to the key properties 
(${\partial_{\theta}}^2 = {\partial_{\bar\theta}}^2 = 0,\,
\partial_{\theta} \,\partial_{\bar\theta}\, + \,\partial_{\bar\theta} \,\,\partial_{\theta} = 0 $)  of the translational generators
(${\partial_{\theta}},\, {\partial_{\bar\theta}} $) along the Grassmanian directions ($\theta,\, \bar\theta$) 
of the supermanifold on which our theory is considered. Furthermore, we note that the nilpotency 
and anticommutativity properties are intertwined together in a beautiful  fashion within the framework 
of the augmented version of BT-superfield formalism where one property depends on the other
and {\it vic{\`e}-versa}.

\noindent
\section{(Anti-)dual BRST Symmetries: A Brief Synopsis}

Ihe (anti-)BRST invariant Lagrangian density (15) respects another set of off-shell
nilpotent ($s_{(a)d}^2 = 0 $) and absolutely anticommutating 
($s_d \, s_{ad} + s_{ad} \,s_{d} = 0$) symmetry transformations ($s_{(a)d}$). 
These transformations are called as the (anti-)dual BRST [or (anti-)co-BRST] symmetry transformations because 
the gauge fixing term\footnote{It can be checked that the 1D co-exterior derivative $\delta = +\,\ast\,\,d\, \ast $
(where $d\, = dt\, \partial_t$) operating  on the 1-form connection $A^{(1)} = dt\,\lambda (x)$ produces $\dot\lambda (x)$ 
(because 
$\delta \,A^{(1)}\,=\,+\,\ast\,\,d\, \ast \, A^{(1)} =\dot\lambda (x)$ is a 0-form). One can add/subtract
other 0-forms ($\phi,\,v $) fields to it as has been done in $(\dot\lambda - \phi - v)$.} 
($ \dot\lambda - v - \phi$) remains invariant under it. These transformations 
are listed as:
\begin{eqnarray}
&& s_{ad}\, \lambda = C,\qquad \qquad s_{ad}\, \phi = \frac{1}{2}\,\dot{C},\qquad \qquad s_{ad}\, v = \,\frac{1}{2}\,{\dot C},
\qquad\qquad s_{ad} \,C = 0, \nonumber\\
&&s_{ad}\, \bar C = \, \frac{i}{2}\,(\dot\phi -\dot v + v' - \phi'),\qquad \qquad s_{ad}\,(\dot\phi -\dot v + v' - \phi') = 0,\nonumber\\
&& s_d\, \lambda = \bar C,\,\qquad \qquad s_d\, \phi = \frac{1}{2}\,\dot{\bar C},\,\,\,\qquad \qquad s_d \,v = \frac{1}{2}\,\dot{\bar C},
\qquad \qquad \quad s_d\, \bar C = 0,\nonumber\\
&& s_d C =   -\,\frac{i}{2}\,(\dot\phi -\dot v + v' - \phi'), \qquad \qquad s_d\,(\dot\phi -\dot v + v' - \phi') = 0,
\end{eqnarray}
where $s_{(a)d}$ are the (anti-)dual BRST symmetry transformations.
One can explicitly check 
that the Lagrangian density (15) transforms to the total time derivatives as:
\begin{eqnarray}
&&s_d\,{\cal L}^{(0)}_b = \frac{\partial}{\partial t}\, \Bigl[\,\frac{\dot{\bar C}}{2}\,(\dot\phi -\dot v + v' - \phi')\Bigr],\quad \quad
s_{ad}\,{\cal L}^{(0)}_b = \frac{\partial}{\partial t} \,\Bigl[\,\frac{\dot C}{2}\,(\dot\phi -\dot v + v' - \phi')\Bigr].
\end{eqnarray}
The above observations demonstrate that the action integral $ S = \int dx \int dt \,{\cal L}_b $ remains invariant 
under the transformations $s_{(a)d}$ for the physically meaningful fields that vanish off at infinity.
According to Noether's theorem, the above continuous symmetry transformations lead to the derivation of the conserved charges as
\begin{eqnarray}
&&Q_{ad} = \int dx\Bigl[ \frac{\dot{C}}{2} (\dot\phi -\dot v + v' - \phi') - C (\dot\lambda - v - \phi)\,\Bigr],\nonumber\\
&&Q_d =\int dx\Bigl[ \frac{\dot{\bar C}}{2} (\dot\phi -\dot v + v' - \phi') - \bar C (\dot\lambda - v - \phi ) \Bigr].
\end{eqnarray}
These charges are found to be nilpotent of order two (i.e. $ Q_{(a)d}^2 = 0$) and they are also 
absolutely anticommutating ($Q_d\,Q_{ad} + Q_{ad}\,Q_d = 0$) in nature where the off-shoot 
$(\ddot \phi - \ddot v + \dot v^\prime - \dot \phi^\prime) + 2 \, (v + \phi - \dot \lambda) = 0$ 
of the EL equations of motion is used\footnote{The above charges (41) can be also expressed as:
$Q_d =\int dx\,\bigl [\frac{\dot{\bar C}}{2}\; (\dot\phi -\dot v + v' - \phi')\, - \frac{\bar C}{2} \;
(\ddot \phi - \ddot v + \dot v^\prime - \dot \phi^\prime \bigr ], 
Q_{ad} = \int dx\,\bigl[ \frac{\dot{C}}{2}\; (\dot\phi -\dot v + v' - \phi')\, - \frac{C}{2} \;
(\ddot \phi - \ddot v + \dot v^\prime - \dot \phi^\prime \,\bigr]$ where we have to use 
$(\ddot \phi - \ddot v + \dot v^\prime - \dot \phi^\prime) + 2 \, (v + \phi - \dot \lambda) = 0$
which emereges from the EL  equations of motion that are derived from the Lagrangian density (15).
From these expressions, the anticommutativity property of $Q_{(a)d} $ becomes obvious and trivial
because we can see explicitly (from the computations of the  of l.h.s. of $s_{ad} Q_d = - i\, \{Q_{ad}, Q_d \} = 0 $ and
$s_{d} Q_{ad} = - i\, \{Q_{d}, Q_{ad} \} = 0$) that it is true because the r.h.s. of both these relations
turn out to be zero
(due to the transformation properties of $Q_{(a)d}$ under the off-shell nilpotent (anti-)co-BRST transformations (39)).}.
This can be verified from $s_d Q_{ad} = - i \; \{ Q_{ad}, Q_d \} = 0$ and 
$s_{ad} Q_{d} = - i \; \{ Q_{d}, Q_{ad} \} = 0$ by computing
the l.h.s. of these relations by exploiting (39) and (41). We have discussed about this issue
in our Appendix A where we have differentiated between the (anti-)BRST and (anti-)co-BRST symmetries of
our present system and have pointed out some {\it novel} features.

With the ghost charge $Q_g = -\,i\,\,\int dx\, \bigl[ C\,\dot{\bar C} + \bar C\,\dot C \bigr]$, the off-shell nilpotent and absolutely
anticommutating (anti-)co-BRST charges obey the following algebra:
\begin{eqnarray} 
&&Q_d ^2 = 0, \qquad\qquad\qquad Q_{ad} ^2 = 0,\qquad\qquad\qquad \{Q_ d,\,Q_{ad}\} = 0,\nonumber\\
&&i\,\bigl[Q_g,\, Q_d\bigr] = -\,Q_d, \qquad\qquad\qquad i\,\bigl[Q_g,\, Q_{ad}\bigr] = +\,Q_{ad}.
\end{eqnarray}
which demonstrates that the ghost numbers of the (anti-)co-BRST charges are $+1$ and $-1$ respectively. Finally, 
we note that the conserved (anti-)co-BRST charges are the generators of continuous symmetry transformations listed in (39).
In the forthcoming two sections, we shall derive the (anti-)co-BRST symmetry transformations and prove the invariance of the 
Lagrangian density in the language of superfield formalism (which captures the nilpotency as well as anticommutativity 
in a simple and straightforward manner).

\noindent
\section{(Anti-)co-BRST Symmetries: Superfield Approach}

To derive the (anti-)co-BRST symmetries (39), we shall exploit the key ingredients of the augmented version of 
superfield approach to BRST formalism [12-15] where we shall demand that all the (anti-)co-BRST invariant 
quantities should remain independent of the ``soul'' coordinates $\theta $ and $\bar\theta$ when the 
former are generalized  onto the (2, 2)-dimensional supermanifold. We observe that 
$s_{(a)d}\,[\phi - v] = 0$. Thus, we demand that (cf. (20)):
\begin{eqnarray}
\tilde\Phi (x, \theta, \bar\theta) - \tilde V (x, \theta, \bar\theta) = \phi(x) - v (x),
\end{eqnarray}
which immediately implies that 
\begin{eqnarray}
\bar f_1 = \bar f_2\equiv \bar f, \qquad\qquad\qquad f_1 = f_2\equiv f, \qquad\qquad\qquad b_1 = b_2 \equiv b.
\end{eqnarray}
We further note that $s_{(a)d} \,[ \dot\lambda - \phi - v] = 0$. This is due to the fact that the 
gauge-fixing term remains invariant under the (anti-)co-BRST symmetry transformations.
Thus, we require the following restrictions on the (super)fields (cf. (20)):
\begin{eqnarray}
\dot{\tilde\lambda}(x, \theta, \bar\theta) - \tilde\Phi (x, \theta, \bar\theta) - \tilde V (x, \theta, \bar\theta)
= \dot\lambda (x) - \phi(x) - v (x).
\end{eqnarray}
Taking the inputs from (44), we obtain the following relationships:
\begin{eqnarray}
\dot{\bar R} = 2\,i\, \bar f, \qquad\qquad\qquad \dot{R} = 2\,i\, f, \qquad\qquad\qquad \dot S = 2\,b.
\end{eqnarray}
Furthermore, the relations (44) also imply the following relationships:
\begin{eqnarray}
\dot{\tilde\Phi} (x, \theta, \bar\theta) - \dot{\tilde V} (x, \theta, \bar\theta) + {\tilde V}^\prime (x, \theta, \bar\theta)
 -\,{\tilde \Phi}^\prime (x, \theta, \bar\theta) = \dot\phi(x) - \dot v (x) + v^\prime (x)- \phi^\prime (x),
\end{eqnarray}
which is true because of the fact that  $s_{(a)d} \,[ \dot\phi - \dot v + v^\prime  - \phi^\prime] = 0$.
Thus, the equality in (47) is due to our basic ideology of the augmented version of superfield formalism [12-15].

Now, we are at a stage where we can exploit the mathematical potential of the (super) co-exterior derivatives
in the application of the dual-horizontality condition because we observe that $s_{(a)d}\,[\dot\lambda - 2\, \phi] = 0$ 
and/or $s_{(a)d}\,[\dot\lambda - 2\,v] = 0$. However, we have also seen that in ordinary 1D space, we have 
$\delta A^{(1)} = \ast\, d\, \ast\, [dt \,\lambda(t)] = \dot\lambda (t) $. Thus, we demand the following
restriction under the dual-horizontality condition:
\begin{eqnarray}
\tilde\delta\, {\tilde A}^{(1)} - 2\, \tilde\Phi (x, \theta, \bar\theta) = \delta \,A^{(1)} -\,2\, \phi (x).
\end{eqnarray}
In our Appendix B, we have explicitly computed the l.h.s and r.h.s where $\tilde\delta =\, \star\, \tilde d\, \star $,
$\tilde d = dt\,\partial_t\, +\, d\,\theta \,\partial_\theta\, + d\,\bar\theta \,\partial_{\bar\theta} $ and $\star $ is the Hodge duality 
operation on the (1, 2)-dimensional super-submanifold (of the general (2, 2)-dimensional supermanifold) under consideration. Ultimately,
our computations yield the following:
\begin{eqnarray}
&&\partial_\theta\, F \,= \,0\qquad \Longrightarrow \qquad \bar B_1 \,= \,0,\qquad s \,= \,0,\nonumber\\
&&\partial_{\bar\theta}\, \bar F \,= \,0 \qquad \Longrightarrow \qquad B_2 \,=\, 0, \qquad \bar s \,= \,0,
\end{eqnarray}
where we have taken the expansions from (20). The above values lead to the determination of the 
modified form of the superfields $F (x, \theta, \bar\theta)$ and $\bar F (x, \theta, \bar\theta)$ as 
\begin{eqnarray}
F^{(r)} (x, \theta, \bar\theta) \,=\, C(x) \,+\, i\, \bar\theta \, B_1(x),\qquad \qquad 
\bar F^{(r)} (x, \theta, \bar\theta) \,=\, \bar C(x) \,+\, i\,\theta \bar B_2(x), 
\end{eqnarray}
where the superscript $(r)$ stands for the reduced form of the superfields.
The above reduced expansions would be used in our further discussions. Finally, we have also the following equality
(see, Appendix B for details):
\begin{eqnarray}
\dot{\tilde\lambda} + (\partial_{\bar\theta}\,F^{(r)}) + (\partial_{\theta}\, \bar F^{(r)}) - 2\, \tilde\Phi = \dot\lambda - 2\,\phi,
\end{eqnarray}
which, ultimately, leads to the following relationships:
\begin{eqnarray}
B_1 + \bar B_2 = 0 \qquad  \Longrightarrow \qquad B_1 = - {\cal B} = - \bar B_2, 
\end{eqnarray}
and the relations written in our earlier equations (46). The above choice, due to the
dual-horizontality condition, leads to the following:
\begin{eqnarray}
F^{(dh)}(x, \theta, \bar\theta) &=& C(x) +\, \theta\,(0) + \, \bar\theta\, (-\,i\,{\cal B}(x)) + \theta\,\bar\theta\, (0)\nonumber\\ 
&\equiv & C(x) + \theta\, (s_{ad}\, C(x))+ \bar\theta\, (s_d\, C(x)) + \theta\, \bar\theta\, (s_d\, s_{ad}\, C(x)),\nonumber\\
\bar F^{(dh)}(x, \theta, \bar\theta) &=& \bar C(x) + \,\theta\, (+\,i\,{\cal B}(x)) + \bar\theta \,(0) + \theta\,\bar\theta \,(0) \nonumber\\ 
&\equiv & \bar C(x) + \theta \,(s_{ad}\, \bar C(x)) + \bar\theta \,(s_d\, \bar C(x)) + \theta \, \bar\theta \,(s_d\, s_{ad}\, \bar C(x)).
\end{eqnarray}
In explicit terms, we have already obtained the transformation properties of (anti-)ghost fields
$(\bar C)$C under the (anti-)dual-BRST transformations as:
\begin{eqnarray}
&&s_d\, C =  - i\,{\cal B}, \qquad \, s_{ad}\, C = 0,\qquad\,\, s_d\, {\cal B} = 0,\qquad s_d\,s_{ad}\, C = 0,\nonumber\\
&&s_d\, \bar C = 0,\qquad s_{ad}\, \bar C =\,+\,i\,{\cal B}, \qquad s_d\, {\cal B} = 0,\qquad s_d\,s_{ad}\, \bar C = 0,
\end{eqnarray}
where $s_{(a)d} \,{\cal B} = 0$ has been obtained due to the requirement of the absolute nilpotency property (i.e. ${s_{(a)d}}^2 = 0$).
We are free to choose ${\cal B}$ in terms of the basic fields so that the nilpotency and the absolute 
anticommutativity properties could be respected together. In this connection, it is worthwhile to point out that:
\begin{eqnarray}
s_{(a)d}\,\bigl[\dot\phi - \dot v + v^\prime - {\phi}^\prime \bigr] = 0.
\end{eqnarray}
Thus, we choose\footnote{In our Appendix C, we provide a precise proof of this 
choice by using the basic ingredients of augmented superfield formalism where the (anti-)co-BRST invariant quantities
are required to be independent of the Grassmannian variables when they are generalized onto an appropriate supermanifold.}
${\cal B} = +\, \frac{1}{2}\, \bigl[\dot\phi - \dot v + v^\prime - {\phi}^\prime \bigr]$.
In this choice, we have been guided by the fact that field ${\cal B}$ should {\it not} explictly depend on the gauge
field $\lambda (t)$. This implies that we have the following:
\begin{eqnarray}
F^{(dh)}(x, \theta, \bar\theta) &=& C(x) + \theta\, (0) + \bar\theta \,(-\, \frac{i}{2}\, \bigl[\dot\phi - \dot v + v^\prime - {\phi}^\prime \bigr]) 
+ \theta\, \bar\theta\, (0),\nonumber\\
\bar F^{(dh)}(x, \theta, \bar\theta) &=& \bar C(x) + \theta\, (+\, \frac{i}{2}\, \bigl[\dot\phi - \dot v + v^\prime - {\phi}^\prime \bigr]) 
+ \bar\theta \,(0) + \theta \,\bar\theta\, (0),
\end{eqnarray}
where the superscript $(dh)$ on the superfields denotes the expansions obtained after the application of the dual-HC.
In other words, we have derived the following (anti-)co-BRST transformations for the
(anti-)ghost fields of our theory, namely;
\begin{eqnarray}
&& s_d C = -\,\frac{i}{2}\,(\dot\phi - \dot v + v^\prime - {\phi}^\prime), \qquad\quad  s_{ad}\, C = 0,\qquad\quad
s_d\,s_{ad} C = 0,\nonumber\\
&& s_d \bar C = 0,\qquad\quad s_{ad}\,\bar C = +\,\frac{i}{2}\,(\dot\phi - \dot v + v^\prime - {\phi}^\prime ), \qquad\quad
s_d\,s_{ad} \bar C = 0,
\end{eqnarray}
where the nilpotency $s_{(a)d}^2 = 0$ and the absolute anticommutativity property $(s_d\,s_{ad} + s_{ad}\,s_d)\,C = 0$ 
and $(s_d\,s_{ad} + s_{ad}\,s_d)\, \bar C = 0$ are satisfied.

Finally, we note from the symmetry of the Lagrangian density (15) that
 $s_{(a)d}\,\bigl[\lambda(\dot\phi - \dot v + v^\prime - {\phi}^\prime) + 2\,i\,\bar C \, C\bigr] = 0$. This
observation leads to the requirement of the following equality:
\begin{eqnarray}
\tilde\lambda\,\bigl[\dot{\tilde\Phi} - \dot{\tilde V} + {\tilde V}' - {\tilde\phi}' \bigr] + 2\,i\,\bar F^{(dh)}\,\bar F^{(dh)}
= \lambda\,(\dot\phi - \dot v + v^\prime - {\phi}^\prime) + 2\,i\,\bar C \, C.
\end{eqnarray}
Using the expansions in (20), (56) and (47), we obtain the following:
\begin{eqnarray}
\bar R = + \, C,\qquad\quad R = + \,\bar C, \qquad\quad S = -\,\frac{1}{2}\,(\dot\phi - \dot v + v^\prime - {\phi}^\prime).
\end{eqnarray}
Now the comparison with (46) yields the following relationships between the secondary fields and other basic fields, namely;
\begin{eqnarray}
 f = - \frac{i}{2}\; \dot {\bar C}, \qquad\quad \bar f = - \frac{i}{2}
\; \dot C,\qquad\quad b = -\,\frac{1}{4}\,(\ddot\phi - \ddot v + {\dot v}^\prime - {\dot\phi}^\prime).
\end{eqnarray}
The above relationships demonstrate that we  have obtained all the secondary fields of the expansion in (20)
in terms of the basic fields of the 2D ordinary theory. Collected together, these relationships are as follows:
\begin{eqnarray}
&&\bar R = + \, C,\qquad\qquad \quad R = + \,\bar C, \,\qquad\qquad\,\, S = -\,\frac{1}{2}\,(\dot\phi - \dot v + v^\prime - {\phi}^\prime),\nonumber\\
&&f = - \frac{i}{2}\; \dot {\bar C}, \qquad\qquad \bar f = - \frac{i}{2}
\; \dot C,\qquad\qquad b = -\,\frac{1}{4}\,(\ddot\phi - \ddot v + {\dot v}^\prime - {\dot\phi}^\prime).
\end{eqnarray}
The substitution of these values in expansion (20) yields the following explicit expansions:
\begin{eqnarray}
{\tilde\lambda}^{(dh)} (x, \theta, \bar\theta) &=& \lambda(x) + \theta\, (C) + \bar\theta \,(\bar C)
+ \,\theta\, \bar\theta\, \bigl[-\,\frac{i}{2}\,(\dot\phi - \dot v + v^\prime - {\phi}^\prime)\bigr] \nonumber\\
&\equiv&  \lambda(x) + \theta\,(s_{ad}\,\lambda)+ \bar\theta \,(s_{d}\,\lambda) + \theta\, \bar\theta\,\,(s_{d}\,s_{ad}\,\lambda), \nonumber\\
F^{(dh)}(x, \theta, \bar\theta ) &=& C(x) + \theta(0) + \bar\theta\, (-\frac{i}{2}\bigl[\dot\phi - \dot v + v^\prime - {\phi}^\prime\bigr])
+ \theta\,\bar \theta\, (0) \nonumber\\
&\equiv& C(x) + \theta\,(s_{ad}\,C)+ \bar\theta \,(s_{d}\,C) + \theta\, \bar\theta\,\,(s_{d}\,s_{ad}\,C), \nonumber\\
\bar F^{(dh)}(x, \theta, \bar\theta ) &=& \bar C(x) +  \theta\, (\frac{i}{2}\,\bigl[\dot\phi - \dot v + v^\prime - {\phi}^\prime\bigr]) + \bar\theta\, (0)
+ \theta\,\bar\theta\,(0) \nonumber\\
&\equiv& \bar C(x) + \theta\,(s_{ad}\,\bar C)+ \bar\theta \,(s_{d}\,\bar C) + \theta\, \bar\theta\,\,(s_{d}\,s_{ad}\,\bar C), \nonumber\\
\tilde\Phi^{(as)} (x, \theta, \bar\theta) &=& \phi(x) + \theta \,(+\, \frac{\dot C}{2}) + \bar\theta\,(+\, \frac{\dot {\bar C}}{2})
+ \theta\,\bar\theta\,(-\,\frac{i}{4}\,\frac{\partial}{\partial t}\;\bigl[\dot\phi - \dot v + v^\prime - {\phi}^\prime\bigr]) \nonumber\\
&\equiv& \phi(x)  + \theta\,(s_{ad}\,\phi)+ \bar\theta \,(s_{d}\,\phi) + \theta\, \bar\theta\,\,(s_{d}\,s_{ad}\,\phi), \nonumber\\
\tilde V^{(as)} (x, \theta, \bar\theta ) &=& v(x) + \theta \,(+\, \frac{\dot C}{2}) + \bar\theta\,(+\, \frac{\dot{\bar C}}{2})
 + \theta\,\bar\theta\,(-\,\frac{i}{4}\,\frac{\partial}{\partial t}\;\bigl[\dot\phi - \dot v + v^\prime - {\phi}^\prime\bigr])\nonumber\\
&\equiv& v(x)  + \theta\,(s_{ad}\,v)+ \bar\theta \,(s_{d}\,v) + \theta\, \bar\theta\,\,(s_{d}\,s_{ad}\,v),
\end{eqnarray}
where the superscripts $(as)$ and $(dh)$ denote the expansions that have been obtained after the 
applications of the augmented superfield formalism and the dual-horizontality condition, respectively. A close
look at the expansions (62) shows that we have already obtained the (anti-)co-BRST symmetries (39)
for all the relevant fields of the theory.

We wrap up this section with the remark that we have also the analogue of mapping (30) between the 
(anti-)co-BRST symmetry transformations $s_{(a)d}$ and the translational generators along the Grassmanian
directions ($\theta,\,\bar\theta$) of the (2, 2)-dimensional supermanifold. This is succinctly
expressed in the mathematical form as follows: 
\begin{eqnarray}
\frac{\partial}{\partial\theta} \Sigma^{(dh,\, as)} (x, \theta, \bar\theta ) \mid _{\bar\theta = 0}\, = s_{ad}\, \sigma  (x), \qquad\quad
\frac{\partial}{\partial\bar\theta} \Sigma^{(dh, \,as)} (x, \theta, \bar\theta ) \mid _{\theta = 0} \,= s_{d}\, \sigma (x),
\end{eqnarray}
where the superscripts $(dh, as)$ on the generic superfield $\Sigma ^{(dh,as)}(x, \theta, \bar\theta )$ 
denote the expansions (for these fields) that have been obtained after the application of the
{dual-}horizontality and the augmented version of the superfield restrictions and $\sigma  (x)$ denotes the
ordinary generic 2D field of the theory. It is obvious now that the nilpotency and the absolute anticommutativity 
of the (anti-)co-BRST symmetry transformations are encoded in such relations 
(i.e. ${\partial_{\theta}}^2 = {\partial_{\bar \theta}}^2 = 0,\, \partial_{\theta}\,\partial_{\bar \theta} + \partial_{\bar \theta}\,\partial_{\theta} = 0 $)
associated with the translational generators ($\partial_{\theta},\, \partial_{\bar\theta}$)
along the Grassmanian directions ($\theta,\,\bar\theta$) of the (2, 2)-dimensional supermanifold.

\noindent
\section{(Anti-)co-BRST Invariance, Nilpotency and Anticommutativity: Superfield Approach}

n this section, we embark on to provide the geometrical meaning to the (anti-)co-BRST invariance 
of the action integral (cf. (40)) and nilpotency and absolute anticommutativity of the (anti-)co-BRST 
charges (cf. (42)) within the framework of our augmented superfield approach to BRST formalism.
First of all, we focus on the (anti-)co-BRST invariance of the action integral $S = \int d^2x \,{\cal L}^{(0)}_b$
where  ${\cal L}^{(0)}_b$ is the Lagrangian density (15) which respects the (anti-)co-BRST symmetries
(as well as the on-shell nilpotent (anti-)BRST symmetries).

Taking the inputs from (62), it is straightforward to note that the Lagrangian density (15) can be generalized onto the 
(2, 2)-dimensional supermanifold in terms of the superfields obtained after application of the dual-horizontality condition
and (anti-)co-BRST invariant restrictions as (cf. (62)):
\begin{eqnarray}
{\cal L}^{(0)}_b \longrightarrow  \tilde {\cal L}^{(0)}_b &=&\,\frac{1}{2} \bigl[\,\dot{\tilde\Phi}^{(as)}\,\dot{\tilde\Phi}^{(as)}
 - \dot{\tilde V}^{(as)}\,\dot{\tilde V}^{(as)}\bigr] + \dot{\tilde V}^{(as)}\,\bigl[ v' - \phi' \bigr]
+ \tilde\lambda^{(dh)} \bigl[\dot\phi - \dot v + \tilde v - \tilde\phi \bigr]\nonumber\\
&-& \frac{1}{2}\,\bigl[  v' - \phi' \bigr]^2 + B(x) \,\bigl[ \dot{\tilde\lambda}^{(dh)} - {\tilde V^{(as)}} - \tilde\Phi^{(as)} \bigl]
+ \,\frac{B^2(x)}{2}
- \,i\,\dot{\bar F}^{(dh)} \,\dot{F}^{(dh)}\nonumber\\ &+& 2\,i\, {\bar F}^{(dh)}\, {F}^{(dh)}.
\end{eqnarray}
First of all, it is important to note that we have taken:
\begin{eqnarray}
\dot{\tilde\Phi}^{(as)} -  \dot{\tilde V}^{(as)} + \tilde V'^{(as)} - \tilde\Phi'^{(as)} &=& \dot\phi(x) - \dot v(x) 
+  v'(x) -  \phi'(x),\nonumber\\
\tilde V'^{(as)}(x, \theta ,\bar\theta) - \tilde\Phi'^{(as)}(x, \theta,\bar\theta) &=& v'(x) - \phi ' (x),
\end{eqnarray}
due to our observations in the  super-expansions (62). Further, we observe that the Lagrangian density 
(15) has been generalized, in a straightforward manner, onto our chosen (2, 2)-dimensional supermanifold just by replacing 
the ordinary 2D fields by their counterpart superfields that have been obtained after the application of 
the dual-horizontality condition and the (anti-)co-BRST invariant restrictions on the supermanifold.

Now, we are in a position to capture the (anti-)co-BRST invariance (40) in terms of the
geometrical quantities that are defined on the (2, 2)-dimensional supermanifold. In other words, we have the following
observations:
\begin{eqnarray}
&&\frac{\partial}{\partial \theta}\,\bigl[\,\tilde {\cal L}^{(0)}_b \bigr] |_{\,{\bar\theta = 0}} \quad = \quad s_{ad}\,{\cal L}^{(0)}_b 
 \quad \equiv\quad  \frac{\partial}{\partial t} \Bigl[\, \frac{\dot C}{2}\,(\dot\phi 
- \dot v + v' - \phi' )\Bigr ],\nonumber\\ 
&&\frac{\partial}{\partial \bar\theta}\,\bigl[\,\tilde {\cal L}^{(0)}_b \bigr] |_{\,{\theta = 0}}\quad = \quad s_{d}\,{\cal L}^{(0)}_b 
 \quad \equiv\quad  \frac{\partial}{\partial t} \Bigl[\, \frac{\dot {\bar C}}{2}\,(\dot\phi - \dot v +  v' - \phi' )\Bigr],
\end{eqnarray}
which, in the ordinary 2D space, shows the (anti-)co-BRST invariance of the Lagrangian density (15).
This is a consequence of the mapping we have obtained in (63).
Geometrically, one can provide an interpretation for the super-Lagrangian density $\tilde{\cal L}^{(0)}_b$ as 
the sum of composite superfields, derived after the application of the dual-horizontality condition and (anti-)co-BRST 
invariant restrictions, such that its translations along $\theta$ or $\bar{\theta}$ directions of the (2, 2)-dimensional 
supermanifold leads to a total time derivative in the ordinary space. As a consequence, the action integral 
$(S = \int d^2 x\, \tilde{\cal L}^{(0)}_b)$ remains invariant under the nilpotent (anti-)co-BRST symmetry transformations
in the ordinary space for the physically well-defined fields which vanish off at infinity.

At this juncture, we concentrate on the geometrical interpretation of the nilpotency and the absolute 
anticommutativity of the (anti-)co-BRST charges $Q_{(a)d}$ in the language of the superfield approach 
to BRST formalism. It can be checked that the (anti-)co-BRST charges can be written, 
in terms of the superfields (62), in two different ways as: 
\begin{eqnarray}
Q_d &=& \int dx \,\Bigl[\frac{\partial}{\partial{\bar\theta}}\,(i\, \bar F^{(dh)}\,\dot F^{(dh)}
+ i\, F^{(dh)}\,\dot{\bar F}^{(dh)})|_{\theta = 0}\,\Bigr]\nonumber\\
&\equiv & \int dx \Bigl[\int d\bar\theta\,\frac{\partial}{\partial \bar\theta}\,(i \bar F^{(dh)}\dot F^{(dh)} 
+ i\,F^{(dh)}\dot{\bar{F}}^{(dh)})|_{\theta =0}\,\Bigr],\nonumber\\
Q_{ad} &=& \int dx \Bigl[\frac{\partial}{\partial\theta}\,(i \bar F^{(dh)}\, \dot F^{(dh)} 
+ i\, F^{(dh)} \dot{\bar F} ^{(dh)})|_{\bar\theta = 0}\, \Bigr] \nonumber\\
&\equiv & \int dx \Bigl[\int d\theta\frac{\partial}{\partial\theta} (i\bar{F}^{(dh)}\dot F^{(dh)} 
+ i F^{(dh)}\dot{\bar F}^{(dh)})|_{\bar\theta = 0}\Bigr].
\end{eqnarray}
The above expressions can be translated into our ordinary 2D space in terms of 
the nilpotent (anti-)co-BRST symmetry transformations
$s_{(a)d}$ (in view of the mapping (63)) as follows:
\begin{eqnarray}
Q_d = i\, \int dx \,s_d\, \bigl[C\, \dot{\bar C} + \bar C \, \dot C \bigr],\qquad\qquad
Q_{ad} = i\, \int dx \,s_{ad} \,\bigl[C\, \dot{\bar C} + \bar C \, \dot C \bigr],
\end{eqnarray}
which show the nilpotency of (anti-)co-BRST charges due to $s^2_{(a)d} = 0$.
In the language of superfield formalism, it is evident that 
$\partial_ {\bar\theta} \,Q_d = 0$ and $ \partial_{\theta}\, Q_{ad} = 0$ because of the nilpotency associated with 
$\partial^2_{\theta} = 0, \,\partial^2 _{\bar\theta} = 0$. In other words, 
we have $s_d\, Q_d = -i\, \{Q_d,\, Q_d\} = 0,\, s_{ad}\,Q_{ad} = -i\, \{Q_{ad},\, Q_{ad}\} = 0$  
due to the definition of the generators that imply $Q_{(a)d}^2 = 0 $. We, ultimately, 
draw the conclusion that the nilpotency of the conserved
(anti-)co-BRST charges is geometrically connected with the nilpotency property 
($\partial^2_{\theta} = 0,\, \partial^2 _{\bar\theta} = 0$) of the translational generators along
$(\theta,\, \bar\theta)$-directions of our chosen  (2, 2)-dimensional supermanifold.

We are at a stage where we shall now provide the geometrical origin of the absolute 
anticommutativity properties of the transformations $s_{(a)d}$ (i.e. $s_d \,s_{ad} + s_{ad}\,s_d = 0$) 
and nilpotent (anti-)co-BRST charges $Q_{(a)d}$ (i.e. $Q_d\, Q_{ad} + Q_{ad}\,Q_d =0 $). Towards this goal in mind,
we observe that the following is true:
\begin{eqnarray}
Q_{ad} = \int dx\, s_d  \bigl (-i\, \dot C\, C \bigr), \qquad\qquad
Q_{d} = \int dx \,s_{ad} \bigl (+i\, \dot {\bar C}\, \bar C \bigr).
\end{eqnarray}
The above expressions, in the language of the symmetry generators, imply that $s_d\, Q_{ad} = -i\, \{Q_{ad},\, Q_d\} = 0$ and
$s_{ad}\,Q_{d} = -i\, \{Q_{d},\, Q_{ad}\} = 0$ due to the off-shell nilpotency of ($s^2_{(a)d} = 0$) of the 
(anti-)co-BRST transformations. Thus, we note that the anticommutativity of the charges (i.e. $Q_d\, Q_{ad} + Q_{ad}\,Q_d =0 $) is 
primarily connected with the nilpotency of the (anti-)co-BRST symmetry transformations. Due to the mapping (63),
it is obvious that we can express the above charges as:
\begin{eqnarray}
&&Q_{ad}\, = \,+\,i\,\int dx \,\frac{\partial} {\partial{\bar\theta}}\,\Bigl[ \dot{F}^{(dh)}\, F^{(dh)}\Bigr]| _{\theta = 0}\,
\,\equiv\, +\,i\, \int dx\, \Bigl[\int d\bar\theta\,\bigl[ \dot{F}^{(dh)}\, F^{(dh)}\Bigr]|_{\theta = 0},\nonumber\\
&&\hskip 0.2cm Q_{d}\, = \,-\,i\,\int dx \,\frac{\partial} {\partial{\theta}}\,
\Bigl[ \dot{\bar F}^{(dh)}\,\bar F^{(dh)}\Bigr]| _{\bar \theta = 0}\,
\,\equiv\, -\,i\, \int dx\, \Bigl[\int d\theta\,\bigl[ \dot{\bar {F}}^{(dh)}\, {\bar F}^{(dh)}\Bigr]|_{\bar\theta = 0},
\end{eqnarray}
which immediately lead to the observation that $\partial_{\bar\theta}\, Q_{ad} = 0$ and $\partial_{\theta}\, Q_{d} = 0$. 
Thus, we note that the absolute anticommutativity of the fermionic (anti-)co-BRST charges (i.e. $\{Q_{d},\, Q_{ad}\} = 0$) 
is connected with the nilpotency (i.e. $\partial^2_{\theta} = 0, \,\partial^2 _{\bar\theta} = 0$) associated with the 
translational generators ($\partial_{\theta}, \,\partial_{\bar\theta}$) along the Grassmanian directions of the 
(2, 2)-dimensional supermanifold.

There are other ways, too, for the (anti-)co-BRST symmetry to be expressed in terms of  the geometrical quantities
on the (2, 2)-dimensional super-manifold. These are as follows:
\begin{eqnarray}
&&Q_{ad} =\int dx \,\Bigl[\frac{\partial}{\partial{\bar\theta}} \,\frac{\partial}{\partial \theta} 
\bigl (i \,\tilde\lambda^{(dh)}\, \dot F^{(dh)} \bigr )\Bigr],\quad\quad\,
Q_{d} = \int dx\,\Bigl[\frac{\partial}{\partial{\bar\theta}} \,\frac{\partial}{\partial \theta}\, 
\bigl (i\, \tilde\lambda^{(dh)}\,\, \dot{\bar F}^{(dh)} \bigr )\Bigr],\nonumber\\
&& Q_{ad} =\int dx\, \Bigl[\frac{\partial}{\partial{\bar\theta}} \,
\frac{\partial}{\partial \theta} \bigl (-\,2\,i\,{\tilde\Phi}^{(as)}\, F^{(dh)} \bigr )\Bigr],\,\,\,
Q_{d} = \int dx\,\Bigl[\frac{\partial}{\partial{\bar\theta}} 
\,\frac{\partial}{\partial \theta}\bigl  (-2\,\,i\,{\tilde {\Phi}}^{(as)}\,\bar F^{(dh)} \bigr )\Bigr],\nonumber\\
&& Q_{ad} =\int dx \,\Bigl[\frac{\partial}{\partial{\bar\theta}} \,\frac{\partial}{\partial \theta} 
\bigl (-2\,\,i\,{\tilde V}^{(as)}\,F^{(dh)} \bigr ) \Bigr],\,\,\,
Q_{d} =\int dx \,\Bigl[\frac{\partial}{\partial{\bar\theta}} \,\frac{\partial}{\partial \theta}
\bigl  (- \,2\,i\,{\tilde {V}}^{(as)}\,\bar F^{(dh)} \bigr )\Bigr].
\end{eqnarray}
Due to the mapping (63), it is evident that we can express the
above expansions in the ordinary 2D space of our BRST invariant gauge theory as:
\begin{eqnarray}
&&Q_{ad} = \int dx\,\Bigl[ s_d\, s_{ad} (i \lambda\, \dot{{C}})\Bigr],\qquad \qquad\,
Q_d = \int dx\,\Bigl[ s_d\, s_{ad}\, (i \lambda\, \dot{\bar C})\Bigr],\nonumber\\
&&Q_{ad} = \int dx\,\Bigl[ s_d\, s_{ad}\, (-2\,i \phi\,{C})\Bigr],\qquad\,\,
Q_d = \int dx\,\Bigl[ s_d\, s_{ad}\, (-2\,i \phi\,\bar{C})\Bigr],\nonumber\\
&&Q_{ad} = \int dx\, \Bigl[s_d\, s_{ad}\, (-\,2\,i\, v \,{C})\Bigr],\qquad
Q_d = \int dx\, \Bigl[s_d\, s_{ad}\, (- 2\,i\,v \,\bar{C})\Bigr].
\end{eqnarray}
By exploiting the nilpotency $s^2_{(a)d} = 0$ and anticommutativity $( s_d\,s_{ad} + s_{ad}\, s_d = 0) $ of the
(anti-)co-BRST symmetry transformations  $s_{a(d)}$, it is clear that we can prove the nilpotency and the 
absolute anticommutativity of the (anti-)co-BRST charges $Q_{a(d)}$ by exploiting the following definitions of the symmetry generators:
\begin{eqnarray}
&& s_d\, Q_d = -i \{ Q_d, Q_{ad} \} = 0,\qquad \qquad\, s_d\, Q_{ad} = -i \{Q_d,Q_{ad} \} = 0, \nonumber\\
&& s_{ad}Q_d = -i \{ Q_d, Q_{ad}\} = 0, \qquad\qquad s_{ad}\, Q_{ad} = -i \{Q_{ad}, Q_{ad} \} = 0.
\end{eqnarray}
If we look at the above charges in terms of superfields and the translational generators 
$( \partial_{\theta},\, \partial_{\bar{\theta}}) $ (cf.(71)), it becomes crystal clear that
 the nilpotency and absolute anticommutativity of the (anti-)co-BRST charges (and corresponding 
transformations) are very intimately connected with such kind of properties associated with 
the translational generators $\partial_{\theta}$ and $\partial_{\bar{\theta}}$. In other words, the properties $(\partial^2_{\theta}= 0,\, \partial^2_{\bar{\theta}} = 0,\, 
\partial_\theta \,\partial_{\bar\theta} + \partial_{\bar\theta}\, \partial_\theta = 0)$ are intertwined 
with the algebraic structures of the (anti-)co-BRST symmetry operators $s_{a(d)}$ and the corresponding conserved and nilpotent charges
$Q_{a(d)}$.

\noindent
\section{Conclusions}

The central results of our present endeavor are the 
precise derivations of the proper (i.e. nilpotent and absolutely anticommuting)
(anti-)BRST and (anti-)co-BRST transformations for the 2D effective Lagrangian density (cf. (15)) 
of a chiral bosonic system at the quantum level. We have provided the geometrical basis for the above nilpotent symmetries
in the language of the nilpotent (i.e. $\partial_\theta^2 = 0, \, \partial_{\bar\theta}^2 = 0$)
translational generators ($\partial_{\theta},\,\partial_{\bar\theta}$) along the 
Grassmanian directions ($\theta, \bar\theta$)
of our chosen (2, 2)-dimensional supermanifold (parameterized by $Z^M = (x^\mu, \theta, \bar\theta)$)
on which our present 2D ordinary field theory has been generalized (within the framework of our augmented 
version of BT-superfield formalism which is geometrically quite intuitive).

The dynamics of our present 2D theory is such that, {\it only} a single component (cf. (4), (18)) of the 2D gauge 
field couples with the self-dual chiral bosonic field and its orthogonal component remains {\it inert} and does not participate 
in the dynamics of our present theory
in any significant manner. This is a {\it novel}  feature of a gauge theory, which, we have never come across 
in our earlier discussions on the $p$-form ($p = 1, 2, 3$) gauge theories [21-25] within the framework of superfield and BRST formalisms.
We have been able to capture this novel feature in the augmented version of our superfield approach 
to BRST formalism and shown that the super-subspace variables ($t, \theta, \bar\theta$), parameterizing
the (1, 2)-dimensional super-submanifold, are good enough
to capture the whole dynamics of the theory (despite the fact that the 2D ordinary theory is considered on
the {\it full} (2, 2)-dimensional supermanifold parameterized by  the superspace variables $Z^M \equiv  (x^\mu, \theta, \bar\theta)$).

The other {\it novel} feature of our present theory is the observation that the off-shell as well as 
on-shell nilpotent (anti-)BRST symmetries exist for the theory but {\it only} the off-shell and absolutely anticommuting
(anti-)co-BRST symmetries are found to exist for  the Lagrangian density (15) (which primarily respects  {\it only} the 
{\it on-shell} nilpotent (anti-)BRST symmetry transformations). It turns out that the off-shell
nilpotent (anti-)co-BRST symmetry transformations (39) are absolutely anticommutating, too.
However, the corresponding  (anti-)co-BRST charges $Q_{(a)d}$ are found to be absolutely anticommutating {\it only} when 
the off-shoot (cf. (75)) of the EL equations of motion (76) (emerging from the Lagrangian 
density (15)) is utilized for its proof (see Appendix B for details).

It was a challenging task for us to check the sanctity for our proposal for the Hodge duality ($\star $) 
operation on a (1, 2)-dimensional super-submanifold (of the full (2, 2)-dimensional supermanifold on which our effective 2D 
theory has been generalized). 
We have accomplished this goal (in our Sec. 6) where we have exploited the Hodge-duality operation in the 
dual-horizontality condition (cf. (79) Appendix B) defined on the (1, 2)-dimensional super-submanifold 
and derived the analogue of the Curci-Ferrari restrictions 
(cf. (52)). This exercise has enabled us to derive the correct off-shell nilpotent (anti-)co-BRST symmetry
transformations of our theory.
The idea of the dual-horizontality condition is  a  {\it new} ingredient in the realm of superfield approach to 
BRST formalism (which we have appropriately applied in the context of our present 2D theory).

We plan to apply the theoretical potential and power of the augmented version of our superfield formalism
to supersymmetric gauge theories of phenomenological interest and establish a connection between the BRST 
type symmetries and supersymmetries.  In particular, we wish to apply this method to 
obtain the precise (anti-)BRST symmetry transformations for the supersymmetric gauge theories.  We have 
already obtained the supersymmetrization of the HC for a SUSY system of spinning relativistic particle
and have obtained its proper (anti-)BRST symmetries and CF-type condition [26]. This direction of investigation
is one of the open problems for us for our future endeavors.\\

\noindent
{\bf Acknowledgements:} TB is grateful to BHU-fellowship and DS thanks UGC,
Government of India, New Delhi, for financial support through RFSMS scheme under which the present investigation
has been carried out. Fruitful comments by S. Krishna 
on some important issues of our present paper are thankfully acknowledged.
We thank the reviewer for his enlightening comments.\\

\begin{center}
\Large{\bf Appendix A: On the (anti-)BRST and (anti-)co-BRST symmetry transformations}\\
\end{center}
\vskip 0.5cm
One of the key observations of our present investigation is the fact that the Lagrangian density (5)
and (15) respect {\it off-shell} and {\it on-shell} nilpotent (anti-)BRST symmetries as quoted in equations (6) and (13),
respectively. On the other hand, the Lagrangian density (15) respects the {\it off-shell} nilpotent (anti-)co-BRST symmetry transformations
 and there is no {\it on-shell} version of it. One of the characteristic and decisive features of this  (anti-)co-BRST 
symmetry transformations is the requirement that the gauge-fixing term ($\dot\lambda - \phi - v$), which 
basically owes its origin to the co-exterior derivative $(\delta\, =\,+ \,\ast \,\, d \, \ast )$, remains invariant under
the {\it above} infinitesimal, off-shell and absolutely anticommutating symmetry transformations.

The interesting properties, that are very important for us, are the nilpotency and anticommutativity.
It is straightforward to note that, for the (anti-)BRST symmetry transformations, the above sacrosanct properties
are satisfied for the off-shell as well as on-shell cases. This statement can be readily  corroborated by the 
equations (10) where the l.h.s. of these relations can be computed by taking the help of (6) and (9) for the
off-shell nilpotent (anti-)BRST symmetries. Similar exercise could be {\it also} performed for the on-shell nilpotent version. This,
however, is not the case with the (anti-)co-BRST symmetries and their corresponding charges. In particular,
in the case of the latter, it can be checked explicitly that the absolute anticommutativity of the (anti-)co-BRST charges.
\begin{eqnarray}
&&s_d\,Q_{ad} = -\,i\,\{ Q_{ad},\, Q_d\}\, =\,0 ,\qquad\qquad
s_{ad}\,Q_{d} = -\,i\,\{ Q_{d},\, Q_{ad}\}\,=\,0,
\end{eqnarray}
can be true if and only if the following equation:
\begin{eqnarray}
\ddot \phi - \ddot v + {\dot v}^\prime - {\dot \phi}^\prime\, = \,2\,(\dot\lambda - v - \phi),
\end{eqnarray}
is satisfied which emerges from the appropriate combination of the  following equations of motion that are derived 
from (15); namely;
\begin{eqnarray}
&&\ddot \lambda = 2 \dot v + {\phi}^\prime - {v}^\prime ,\nonumber\\
&&\ddot v - 2\,\dot v' + {\dot\phi}' + (\dot\lambda  - \lambda' ) -({\phi}'' - v'') + (\dot\lambda - v - \phi) = 0,\nonumber\\ 
&&\ddot\phi + (\dot\lambda  - \lambda' ) - \dot v' - (\phi'' - v '') - (\dot\lambda - v - \phi) = 0.  
\end{eqnarray}
We would like to state, in passing, that
the relation of type (75) is the analogue of the celebrated Curci-Ferrari conditions for 
this system which turns out to be an (anti-)co-BRST invariant quantity.  In other words, we have 
$s_{(a)d} \,\bigl[(\ddot \phi - \ddot v + \dot v^\prime - \dot \phi^\prime)
- 2 \, ( \dot \lambda - v - \phi)\bigr] = 0$.

We close this Appendix with the remark that, within the framework of superfield approach to BRST formalism, 
too, there is a key difference between the nilpotent (anti-)BRST and (anti-)co-BRST symmetries. In the former case, we have 
$\partial_{\theta}\, \bar F = 0$ and $\partial_{\bar\theta}\,F = 0$ when we set the coefficients of $(d\theta \wedge d\theta)$ 
and $(d\bar\theta \wedge d\bar\theta)$ equal to zero in the HC (i.e. $\tilde d\, {\tilde A}^{(1)} = d \,A^{(1)} $)
which leads to the relationships (23). On the other hand, for the case of the latter, we obtain the conditions
$\partial_{\theta}\,F = 0$ and $\partial_{\bar\theta}\,\bar F = 0$ which lead to the relations (49). The CF-type 
restrictions that emerge from this (dual-)horizontality conditions are (cf. (23),(52)):
\begin{eqnarray}
\bar B_1 +  B_2 = 0, \qquad\qquad\qquad  B_1 + \bar B_2 = 0,
\end{eqnarray}
which enable us to obtain the absolute anticommutativity of the 
nilpotent (anti-)co-BRST and (anti-)BRST symmetry transformations, respectively.\\

\begin{center}
\large{\bf Appendix B: Hodge duality operation on the (1, 2)-dimensional super-submanifold}\\
\end{center}
\vskip 0.5cm
In this Appendix, we demonstrate, that the action of the Hodge-duality operation on a
(1, 2)-dimensional super-submanifold (of the full (2, 2)-dimensional supermanifold) in the following
 (super) 0-form relationship due to the dual-horizontality condition, namely;
\begin{eqnarray}
\tilde\delta\, {\tilde A}^{(1)} - 2\, \tilde\Phi (x, \theta, \bar\theta) = \delta \,A^{(1)} - 2\,\phi (x).
\end{eqnarray}
where $\delta = \ast\, \, d\, \ast $ and  $\tilde\delta = \star\,\, \tilde d\, \star $. 
Here the operator $\ast $ is the Hodge-duality operation on the 1D ordinary sub-manifold
of the full 2D ordinary spacetime manifold (parameterized by $t$) and $\star $ is the Hodge-duality 
operation on the (1, 2)-dimensional super-submanifold characterized by the supervariables $(t, \theta, \bar\theta)$.
It can be seen that the $\tilde\delta\, \tilde A^{(1)}\,=\,+\, \star\, \tilde d\, \star \, \tilde A^{(1)} $ can 
be computed from (19) in the following manner. First of all, we take the operation $(\star\,\tilde A^{(1)} )$
as given below in an explicit manner:
\begin{eqnarray}
\star\,\tilde A^{(1)} &=& \, \star\, \bigl[ dt \,\tilde\lambda (x, \theta, \bar\theta) +\, d\theta \bar F(x, \theta, \bar\theta) 
+\, d\,\bar\theta\, F (x, \theta, \bar\theta )\bigr]\nonumber\\
&=& (d\theta \,\wedge\,d\bar\theta )\, \tilde\lambda (x, \theta, \bar\theta)\,+\, (dt\, \wedge\,d\bar\theta ) \,\bar F(x, \theta, \bar\theta)
+\, (dt\, \wedge\,d\theta)\,F(x, \theta, \bar\theta),
\end{eqnarray}
which is nothing but a super 2-form on the (1, 2)-dimensional superfield. Now the application $\tilde d$ on (79) leads 
to the following super 3-form:
\begin{eqnarray}
\tilde d\, (\star\,\tilde A^{(1)}) &=&(dt\,\wedge\,d\theta \,\wedge\,d\bar\theta )\, \dot{\tilde\lambda} 
+ (dt\,\wedge\,dt \, \wedge\,d\bar\theta ) \,\dot{\bar F}
+ (dt\, \wedge\,dt\,\wedge \, d\,\theta)\,\dot F\nonumber\\ &+& (d\theta\,\wedge\,d\theta \,\wedge\,d\bar\theta)\, \partial_{\theta}\,\tilde\lambda
-(d\theta\,\wedge\,dt \,\wedge\,d\bar\theta)\,\partial_{\theta}\, \bar F -(d\theta\,\wedge\,dt \,\wedge\,d\theta)\,\partial_{\theta}\,F\nonumber\\
&+& (d\bar\theta\,\wedge\,d\theta \,\wedge\,d\bar\theta)\,\partial_{\bar\theta}\,\tilde\lambda
-(d\bar\theta\,\wedge\,dt \,\wedge\,d\bar\theta)\,\partial_{\bar\theta}\,\bar F
-(d\bar\theta\,\wedge\,dt \,\wedge\,d\theta)\,\partial_{\bar\theta}\,F.
\end{eqnarray}
In the above, the second and the third terms in the {\it first} line should be zero because 
$(dt\,\wedge\,dt) = 0$. The first terms of the second and the third lines would be zero 
according to the prescription laid down in [27] because, on a (1,  2)-dimensional superfield,
 a super 3-form with {\it merely} ($\theta, \bar\theta$) terms {\it cannot} be defined 
(see, e.g. [27] for details).
Thus, ultimately, we have the following:
\begin{eqnarray}
\tilde d\, (\star\,\tilde A^{(1)}) &=& (dt\,\wedge\,d\theta \,\wedge\,d\bar\theta )\, \dot{\tilde\lambda} 
+ (dt\,\wedge\,d\theta \,\wedge\,d\bar\theta)\,\partial_{\theta}\, \bar F + (dt\,\wedge\,d\theta \,\wedge\,d\theta)\,\partial_{\theta}\,F\nonumber\\
&+& (dt\,\wedge\,d\bar\theta \,\wedge\,d\bar\theta)\,\partial_{\bar\theta}\,\bar F 
+ (dt \,\wedge\,d\theta\,\wedge\,d\bar\theta\,)\,\partial_{\bar\theta}\,F,
\end{eqnarray}
where we have used $(dt \,\wedge\,d\theta) = - (d\theta\,\wedge\,dt)$ and $(dt \,\wedge\,d\bar\theta) = - (d\bar\theta\,\wedge\,dt)$.
Now, we are at a stage, to apply the final $\star $ on (81). Using the rules of [27], we have the following operation
on the 3-form differentials of the (1, 2)-dimensional super-submanifold:
\begin{eqnarray}
\star \,(dt\,\wedge\,d\theta \,\wedge\,d\bar\theta ) = 1, \qquad \star\, (dt\,\wedge\,d\theta \,\wedge\,d\theta) = S^{\theta\theta},\qquad
\star \,(dt\,\wedge\,d\bar\theta \,\wedge\,d\bar\theta) = S^{\bar\theta \bar\theta}, 
\end{eqnarray}
where $S^{\theta\theta}$ and $S^{\bar\theta \bar\theta}$ are symmetric so that if we take another $\star $ on  (82) 
we should get back the original 3-form differentials (modulo a $\pm$ sign). It is evident that all the super 0-forms in (82)
are independent of one-another. 

Taking the above into account, we have the following equation, namely;
\begin{eqnarray}
\star \, \tilde d \star\,\tilde A^{(1)} = (\dot{\tilde\lambda} + \partial_{\theta} \bar F + \partial_{\bar\theta}\,F) +
 S^{\theta\theta} \partial_{\theta}F + S^{\bar\theta \bar\theta}\partial_{\bar\theta} \bar F.
\end{eqnarray}
It is now obvious that the equality (78) can be expanded as follows:
\begin{eqnarray}
(\dot{\tilde\lambda} + \partial_{\theta}\, \bar F + \partial_{\bar\theta}\,F - 2\, \tilde\Phi) 
+ S^{\theta\theta}\, \partial_{\theta}\,F + S^{\bar\theta \bar\theta}\,\partial_{\bar\theta}\,\bar F 
= \dot{\lambda} - 2\,\phi.
\end{eqnarray}
This is the explicit expression that has been taken in the main body of our text where we have computed the
(anti-)dual BRST symmetry transformations for the (anti-)ghost fields of our theory in a correct fashion (cf. (53)).

\vskip 1.0cm

\begin{center}
\Large{\bf Appendix C: On the specific choice of $\cal {B}$}
\end{center}
\vskip 0.7cm
In this Appendix, we provide the precise derivation of the ad-hoc choice made in Sec. 6 (cf. comment after equation (55))
for the auxiliary variable
\begin{eqnarray}
\cal{B} &=& \frac{1}{2}(\dot{\phi} - \dot{v} + v' - \phi'),
\end{eqnarray}
from which ensues the relations
$ R = \bar{C},\, \bar{R} = C,\, f = -\,\frac{i}{2}\,\dot{\bar{C}},\,
\bar{f} = -\, \frac{i}{2}\dot{C},\, b = -\,\frac{1}{4}(\ddot{\phi}- \ddot{v} + \dot{v}' - \dot{\phi}')$
in the main body of our text (cf. Sec. 6).
To derive these explicitly, we have to take recourse to the augmented version of supervariable approach 
where all the (anti-)BRST [and/or (anti-)co-BRST] invariant quantities (of physical interest) are required to be 
independent of the Grassmanian variables $\theta$ and $\bar{\theta}$ when they are generalized onto an appropriate supermanifold.

According to the above logic, we observe that the following additional quantities ($Q_i, i= 1,2,3,4 $)
of interest, namely; 
\begin{eqnarray}
Q_1 = \lambda\, C,\qquad\qquad Q_2 = \phi \,\dot{C},\qquad\qquad Q_3 = \lambda\, \bar{C},\qquad\qquad Q_4= \phi\, \dot{\bar{C}},
\end{eqnarray}
remain invariant under the nilpotent (anti-)co-BRST symmetry transformations 
$s_{(a)d}$ (i.e. $s_{ad}Q_1 = s_{ad}[\lambda \,C] = 0,\, s_{ad}\,Q_2 =  s_{ad}\,[\phi\, \dot{C}] = 0,
\,s_{d}\,Q_3 = s_d\, [\lambda\, \bar{C}] = 0,\,s_{d}\,Q_4 = s_d \,[\phi\, \dot{\bar{C}}] = 0$).
Thus, we demand that the following equalities should be true, namely;
\begin{eqnarray}
\tilde{\lambda}(x, \theta, \bar{\theta})\, \bar{F}^{(dh)} (x, \theta, \bar{\theta}) &=& \lambda(x)\,\bar{C}(x),\nonumber\\
\tilde{\lambda}(x, \theta, \bar{\theta}) \,F^{(dh)}(x, \theta, \bar{\theta}) &=& \lambda(x) \, C(x),\nonumber\\
\tilde{\Phi}(x, \theta, \bar{\theta}) \,\dot{\bar{F}}^{(dh)}(x, \theta, \bar{\theta}) &=& \phi(x) \, \dot{\bar{C}}(x),\nonumber\\
\tilde{\Phi}(x, \theta, \bar{\theta}) \,\dot{F}^{(dh)}(x, \theta, \bar{\theta}) &=& \phi(x) \, \dot C(x),
\end{eqnarray}
which lead to the following relationships:
\begin{eqnarray}
&& i\,\lambda\,{\cal B} + \bar R\, \bar C = 0,\quad\qquad R\,\bar C = 0,\quad\qquad R\,{\cal B} + S\,\bar C = 0,\nonumber\\
&& i\,\lambda\,{\cal B} -  R\, C = 0,\qquad\quad \bar R\, C = 0,\qquad\quad \bar R\,{\cal B} + S\, C = 0,\nonumber\\
&& \phi\,\dot{\cal B} + \bar f\, \dot{\bar C} = 0,\qquad\quad\,\,\, \bar f \dot C = 0, \qquad\quad\,\,\, i\,b\,\dot{\bar C} - f\,\dot{\cal B} = 0,\nonumber\\
&& \phi\,\dot{\cal B} -  f\, \dot C = 0,\qquad\quad\,\,\, f\dot{\bar C} = 0, \qquad\quad\,\,\, i\,b\,\dot C - \bar f\,\dot{\cal B} = 0.
\end{eqnarray}
The above equations automatically show that we have 
$R \propto \bar{C},\, f\propto \dot{\bar{C}}, \,\bar{R} \propto C$ and $\bar{f} \propto \dot{C}$. We make one
of simplest choices: $R = \bar{C},\, \bar{R} = C$. This choice is good enough to lead to all the values written in (61)(cf. Sec. 6).
Thus, we observe that we have derived all the secondary variables in terms of the basic and auxiliary variables
of the theory which fully lead to the expansions (62) that incorporate all the 
transformations $s_{(a)d}$.

\end{document}